\pdfoutput=1

\documentclass[11pt,twoside,a4paper,cmspaper,final,collab]{cms-tdr}
\def\svnVersion{3fcf09a-D}\def\svnDate{2019/12/04}\def\cmsCernNoTag{CERN-EP-2019-137}\def\cmsCernDate{\today}\def\cmsMessage{"Published in the Journal of High Energy Physics as \href{http://dx.doi.org/10.1007/JHEP12(2019)062}{\doi{10.1007/JHEP12(2019)062}.}"}

\begin{document}\cmsNoteHeader{SMP-18-008}

\hyphenation{had-ron-i-za-tion}
\hyphenation{cal-or-i-me-ter}
\hyphenation{de-vices}

\newcommand{\msd}{\ensuremath{m_{\text{SD}}}\xspace}
\providecommand{\PV}{\ensuremath{\text{V}}\xspace}
\newcommand{\WW}{\PW\PW}
\newcommand{\WZ}{\PW\PZ}
\newcommand{\mwv}{\ensuremath{m_{\PW\PV}}\xspace}
\newcommand{\Wjets}{\ensuremath{\PW\text{+jets}}\xspace}
\newcommand{\taun}{\ensuremath{\tau_{N}}\xspace}
\newcommand{\cwww}{\ensuremath{c_{\PW\PW\PW}}\xspace}
\newcommand{\cw}{\ensuremath{c_{\PW}}\xspace}
\newcommand{\cb}{\ensuremath{c_{\PB}}\xspace}
\newcommand{\ci}{\ensuremath{c_{i}}\xspace}
\newcommand{\cj}{\ensuremath{c_{j}}\xspace}
\newcommand{\lambdaz}{\ensuremath{\lambda_{\PZ}}\xspace}
\newcommand{\deltagz}{\ensuremath{\Delta g_{1}^{\PZ}}\xspace}
\newcommand{\deltakapz}{\ensuremath{\Delta \kappa_{\PZ}}\xspace}
\newcommand{\deltakapg}{\ensuremath{\Delta \kappa_{\Pgg}}\xspace}
\newcommand{\pp}{\Pp\Pp}
\newcommand{\SR}{\ensuremath{\text{SR}}\xspace}
\newcommand{\SB}{\ensuremath{\text{SB}}\xspace}
\newcommand{\MC}{\ensuremath{\text{MC}}\xspace}
\newcommand{\muF}{\ensuremath{\mu_{\mathrm{F}}}\xspace}
\newcommand{\muR}{\ensuremath{\mu_{\mathrm{R}}}\xspace}

\providecommand{\cmsTable}[1]{\resizebox{\textwidth}{!}{#1}}
\newlength\cmsTabSkip\setlength{\cmsTabSkip}{1ex}

\cmsNoteHeader{SMP-18-008}
\title{Search for anomalous triple gauge couplings in $\WW$ and $\WZ$ production in lepton + jet events in proton-proton collisions at $\sqrt{s} = 13\TeV$}

\date{\today}

\abstract{A search is presented for three additional operators that would lead to anomalous $\PW\PW\gamma$ or $\PW\PW\PZ$ couplings with respect to those in the standard model. They are constrained by studying events with two vector bosons; a $\PW$ boson decaying to $\Pe\Pgn$ or $\Pgm\Pgn$, and a $\PW$ or $\PZ$ boson decaying hadronically, reconstructed as a single, massive, large-radius jet.
The search uses a data set of proton-proton collisions at a centre-of-mass energy of 13\TeV, recorded by the CMS experiment at the CERN LHC in 2016, and corresponding to an integrated luminosity of 35.9\fbinv. Using the reconstructed diboson invariant mass, 95\% confidence intervals are obtained for the anomalous coupling parameters of $-1.58 < \cwww/\Lambda^2 < 1.59\TeV^{-2}$, $-2.00 < \cw/\Lambda^2 < 2.65\TeV^{-2}$, and $-8.78 < \cb/\Lambda^2 < 8.54\TeV^{-2}$, in agreement with standard model expectations of zero for each parameter. These are the strictest bounds on these parameters to date.}

\hypersetup{%
pdfauthor={CMS Collaboration},%
pdftitle={Search for anomalous triple gauge couplings in WW and WZ production in lepton + jet events in proton-proton collisions at sqrt(s) = 13 TeV},%
pdfsubject={CMS},%
pdfkeywords={CMS, physics, EFT, aTGC, WW, WZ, WV, jet, substructure}}

\maketitle

\section{Introduction}

The standard model (SM) of particle physics provides a thoroughly tested description of the known elementary particles and their interactions.
Its theoretical and observational shortcomings may be explained by the existence of further inner structure at shorter distances or, equivalently, higher energies. One of the goals of the LHC and its detectors is to reveal such structure if it exists.

\begin{figure}[htb]
\centering
\includegraphics[width=0.5\textwidth]{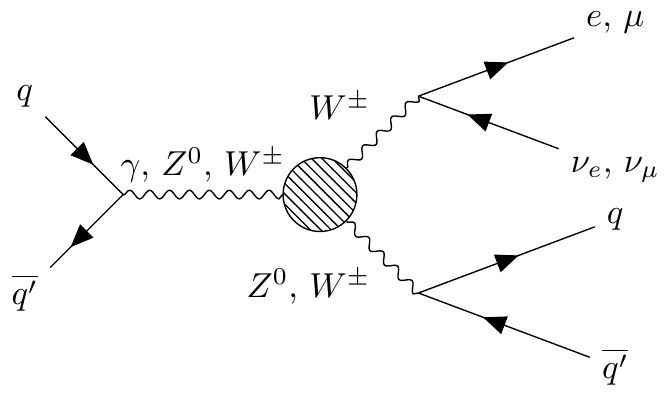}
\caption{The LO Feynman diagram for the diboson process involving triple gauge couplings studied in this analysis. One $\PW$ boson decays to a lepton and a neutrino, and the other $\PW/\PZ$ boson decays to a quark-antiquark pair.}
\label{fig:FeynmanDiagram}
\end{figure}

If the physics beyond the SM does not contain new low-mass particles and is consistent with the symmetries of the SM, its effects can be parametrized in terms of an effective field theory (EFT).
In this approach, the new-physics model is constructed by expanding around the SM and integrating over degrees of freedom at higher energies.
This leads to additional terms in the Lagrangian, proportional to inverse powers of the mass scale of the new particles, up to numerical factors that depend on the new couplings.
We refer to the overall energy scale suppressing these terms as $\Lambda$.
In this paper we focus on possible additional contributions to the production of $\WW$ and $\WZ$ final states parametrized in such an EFT model by dimension-six operators~\cite{PhysRevD.48.2182,Degrande:2012wf}, with the following CP-conserving modification to the SM Lagrangian:
\begin{linenomath}
\begin{equation}
\delta \mathcal{L} = \frac{\cwww}{\Lambda^2}\text{Tr}\left[\PW_{\mu\nu}\PW^{\nu\rho}\PW^{\mu}_{\rho}\right] + \frac{\cw}{\Lambda^2}\left(\text{D}_{\mu} \Phi\right)^{\dag} \PW^{\mu\nu} \left(\text{D}_{\nu} \Phi\right) + \frac{\cb}{\Lambda^2}\left(\text{D}_{\mu} \Phi\right)^{\dag} \PB^{\mu\nu} \left(\text{D}_{\nu} \Phi\right),
\label{eq:atgclagrangian}
\end{equation}
\end{linenomath}
where $\Phi$ is the SM Higgs boson field doublet and
\begin{linenomath}
\begin{equation}
\begin{aligned}
\text{D}_{\mu} &= \partial_{\mu} + \frac{i}{2}g\tau_{I}\PW^{I}_{\mu} + \frac{i}{2}g'\PB_{\mu} \\
\PW_{\mu\nu} &= \frac{i}{2}g\tau_{I} \left(\partial_{\mu}\PW^{I}_{\nu} - \partial_{\nu}\PW^{I}_{\mu} + g\epsilon^{I}_{JK}\PW^{J}_{\mu}\PW^{K}_{\nu}\right) \\
\PB_{\mu\nu} &= \frac{i}{2}g' \left(\partial_{\mu}\PB_{\nu} - \partial_{\nu}\PB_{\mu}\right). \\
\end{aligned}
\end{equation}
\end{linenomath}
The parameters $\{ \cwww, \cw, \cb \}$ control the size of each new contribution.
These additional contributions induce triple gauge couplings (TGCs) beyond those present in the SM, and are referred to as anomalous TGCs (aTGCs).
The SM behaviour is therefore recovered when $\cwww = \cw = \cb = 0$.
Nonzero aTGCs would lead to increased $\WW$ and $\WZ$ production cross sections at high vector boson pair invariant masses.
The search for nonzero aTGCs is performed in the semileptonic final state, with one $\PW$ boson decaying to a lepton ($\Pe$ or $\Pgm$) and a neutrino, and the other $\PW$ or $\PZ$ boson decaying hadronically.
The leading-order (LO) Feynman diagram for this process involving triple gauge couplings is shown in Fig.~\ref{fig:FeynmanDiagram}.

Although the hadronic decay channel of a gauge boson has a larger branching fraction than the leptonic decay channel, it suffers from the presence of background processes with significantly larger cross sections, especially those producing multiple hadronic jets.
The semileptonic final state therefore offers a good balance between efficiency and purity.
It also allows a full kinematic reconstruction of the diboson system, using the $\PW$ mass to constrain the combined four-momentum of the lepton and neutrino.
Since the effects of the aTGCs are most dramatic at high boson momenta, we consider only hadronic decays from highly Lorentz-boosted vector bosons where the hadronization products of the two final state quarks overlap in the detector to form a single, large-radius jet.
This analysis distinguishes $\WW$ and $\WZ$ production using the invariant mass of the jet created as the result of the hadronic decay of the $\PW/\PZ$ boson, thereby providing some discrimination between the different aTGC contributions.
However, the relatively poor jet mass resolution significantly limits this separation power.
Further discrimination between the different aTGC parameters is only possible by studying angular variables that characterize the diboson production and decay products~\cite{Hagiwara:1986vm,Buchalla:2013wpa}.
Such analysis is outside the scope of this search.
To reduce contributions from the significant $\Wjets$ SM background processes, jet substructure techniques are used for the boson identification~\cite{CMS-PAS-JME-16-003}.

Previous searches for such signatures by the ATLAS and CMS experiments have focused on leptonic decays~\cite{Chatrchyan:2012sga,Chatrchyan:2013yaa,CMS:2014xja,Khachatryan:2015sga,Khachatryan:2016poo,Sirunyan:2017zjc,Aad:2011cx,Aad:2011xj,Aad:2012rxl,Aad:2012twa,ATLAS:2012mec,Aad:2012awa,Aad:2016wpd,Aad:2016ett,ATLAS:2012mec,Aad:2016wpd,Aad:2016ett,Aaboud:2016urj,Aaboud:2017rwm,Sirunyan:2019bez,ATLAS-CONF-2016-043,Aaboud:2019nkz}. Earlier studies in the semileptonic final states~\cite{Chatrchyan:2012bd,Sirunyan:2017bey,Aad:2014mda,Aaboud:2017cgf} were performed using data taken at centre-of-mass energies of 7 and 8\TeV.
Similar boosted-boson reconstruction techniques were also used at a centre-of-mass energy of 13\TeV, in the context of search for a narrow resonance decaying to $\WW$ or $\WZ$ in the semileptonic final state~\cite{Sirunyan:2018iff}.

In this paper, the detector is described in Section~\ref{sec:detector}; the data and simulated samples are described in Section~\ref{sec:samples}; the object reconstruction and the event selection are described in Section~\ref{sec:eventSelection}; the signal and background modelling are described in Sections~\ref{sec:signalModel} and~\ref{sec:backgroundModel}, respectively; and the systematic uncertainties affecting this analysis are described in Section~\ref{sec:syst}.
The results are shown in Section~\ref{sec:results}, and a summary is presented in Section~\ref{sec:summary}.

\section{The CMS detector}
\label{sec:detector}

The central feature of the CMS apparatus is a superconducting solenoid of 6\unit{m} internal diameter, providing a magnetic field of 3.8\unit{T}. Within the solenoid volume are a silicon pixel and strip tracker, a lead tungstate crystal electromagnetic calorimeter (ECAL), and a brass and scintillator hadron calorimeter (HCAL), each composed of a barrel and two endcap sections. Forward calorimeters extend the pseudorapidity ($\eta$) coverage provided by the barrel and endcap detectors. Muons are detected in gas-ionization chambers embedded in the steel flux-return yoke outside the solenoid.

Events of interest are selected using a two-tiered trigger system~\cite{CMSTriggerPaper}. The first level, composed of custom hardware processors, uses information from the calorimeters and muon detectors to select events at a rate of around 100\unit{kHz} within a time interval of less than 4\mus. The second level, known as the high-level trigger, consists of a farm of processors running a version of the full event reconstruction software optimized for fast processing, and reduces the event rate to around 1\unit{kHz} before data storage.

A more detailed description of the CMS detector, together with a definition of the coordinate system used and the relevant kinematic variables, can be found in Ref.~\cite{CMSIntroPaper}.

\section{Data and simulated samples}
\label{sec:samples}

The analysis is performed on proton-proton ($\pp$) collision data recorded by the CMS detector in 2016 at a centre-of-mass energy of 13\TeV, corresponding to an integrated luminosity of 35.9\fbinv.

The signal is simulated using \MGvATNLO v2.4.2~\cite{Alwall:2014hca} at next-to-leading order (NLO) in the strong coupling $\alpS$, using the ``EWDim6'' model, which implements the aforementioned EFT~\cite{Degrande:2012wf}.
The simulated signal processes include decays of the $\PW$ boson to a tau lepton and neutrino, with the subsequent decay of the tau lepton to a muon or electron and the accompanying neutrino.
The simulated signal events are first generated with all three aTGC parameters set to nonzero positive values, and then reweighted to different permutations of zero and nonzero aTGCs using the matrix-element event weights computed by \MGvATNLO.
This includes the scenario where all aTGCs parameters are zero, corresponding to SM diboson production.
The signal sample is rescaled such that the cross sections for diboson production in this scenario are normalized to the corresponding SM next-to-next-to-leading order (NNLO) cross sections~\cite{Grazzini:2016swo,Gehrmann:2014fva} as described in Section~\ref{sec:signalModel}.

For the simulation of background SM processes, a variety of event generators are used.
The \POWHEG v1.0~\cite{Nason:2004rx,Frixione:2007vw,Alioli:2010xd,Re:2010bp} generator is used for the generation of $\cPqt\PW$ events, whilst \POWHEG v2.0~\cite{Alioli:2009je,Alioli:2011as,Campbell:2014kua,Frederix:2012dh,Frixione:2007nw,Melia:2011tj,Nason:2013ydw} is used for the generation of \ttbar and $t$-channel single top quark events, all at NLO.
The \MGvATNLO v2.2.2 generator is used to generate \Wjets and $s$-channel single top quark processes at NLO.
The parton showering and hadronization for all samples are performed with \PYTHIA~\cite{Sjostrand:2014zea}, using v8.205 for the $s$-channel single top quark samples, and v8.212 for all other samples.
The FxFx merging scheme~\cite{fxfx,Alwall:2014hca} is used for samples generated at NLO, and the MLM merging scheme~\cite{Alwall:2007fs} for those generated at LO.
The CUETP8M2T4 underlying event tune~\cite{CMS-PAS-TOP-16-021} is used for the \ttbar sample, whilst the CUETP8M1 underlying event tune~\cite{CMS-PAPER-GEN-14-001} is used for all other samples.

The {\Wjets} samples are normalized using inclusive cross sections calculated at NNLO using \MCFM~v6.6~\cite{MCFM:VJets}.
The \textsc{Top++}2.0~\cite{Czakon:2011xx} program is used to calculate the \ttbar cross section at NNLO in quantum chromodynamics (QCD), including resummation of next-to-next-to-leading logarithmic soft gluon terms.

All events are generated with the NNPDF 3.0 parton distribution functions (PDFs)~\cite{Ball:2014uwa}.
Detector response in the Monte Carlo (MC) samples is simulated using a detailed description of the CMS detector implemented with the {\GEANTfour}~\cite{Agostinelli:2002hh} package, and processed using the same software chain used for collision data.
Residual differences between data and simulation with respect to jet energy scale, jet energy resolution, jet {\cPqb} tagging efficiency, lepton identification efficiency, lepton energy scale, trigger efficiency, and jet substructure selection efficiency are corrected by corresponding scale factors.
Minimum-bias events are superimposed on the simulated events to emulate the effects of additional $\pp$ interactions within the same or nearby bunch crossings (pileup), with an average number of 23 $\pp$ collisions per bunch crossing.
All simulated samples are reweighted to match the distribution of the number of $\pp$ interactions per bunch crossing as measured in the data.

\section{Object reconstruction and event selection}
\label{sec:eventSelection}

Events targeting the electronic decay of the $\PW$ boson are selected by a single-electron trigger that requires the event to contain either (i) at least one electron candidate satisfying ``loose'' isolation criteria with transverse momentum $\pt>45\GeV$ and $\abs{\eta}<2.5$, or (ii) at least one electron candidate with $\pt>115\GeV$ and $\abs{\eta}<2.5$ without any additional electron isolation criteria~\cite{Khachatryan:2015hwa,Sirunyan:2018iff}.
For the muonic $\PW$ boson decay channel, data are selected by a single-muon trigger~\cite{Sirunyan:2018} that requires an event to contain at least one muon candidate with $\pt > 50 \GeV$ and $\abs{\eta} < 2.4$.

Events accepted for analysis must pass a number of quality criteria designed to reject events containing significant noise in any of the subdetectors, and are also required to have at least one well-reconstructed collision vertex.
The reconstructed vertex with the largest value of summed object $\pt^2$ is the primary $\pp$ interaction vertex.
The objects considered are (i) jets clustered using the anti-\kt jet algorithm~\cite{Cacciari:2008gp,Cacciari:2011ma}, with the tracks assigned to the vertex as the input, and (ii) the associated missing transverse momentum, taken as the negative vector sum of the \pt of those jets, to account for neutral particles.
More details are given in Section~9.4.1 of Ref.~\cite{CMS-TDR-15-02}.

The particle-flow (PF) algorithm~\cite{Sirunyan:2017ulk} aims to reconstruct and identify each individual particle in an event, with an optimized combination of information from the various elements of the CMS detector. The energy of photons is obtained from the ECAL measurement. The energy of electrons is determined from a combination of the electron momentum at the primary interaction vertex as determined by the tracker, the energy of the corresponding ECAL cluster, and the energy sum of all bremsstrahlung photons spatially compatible with originating from the electron track. The momentum of muons is obtained from the curvature of the corresponding track. The energy of charged hadrons is determined from a combination of their momentum measured in the tracker and the matching ECAL and HCAL energy deposits, corrected for zero-suppression effects and for the response of the calorimeters to hadronic showers. Finally, the energy of neutral hadrons is obtained from the corresponding corrected ECAL and HCAL energies.

Electrons are reconstructed by combining information from the central tracking detector and ECAL~\cite{Chatrchyan:2013dga, Khachatryan:2015hwa}.
Electron candidates are required to exceed a transverse momentum threshold of 50\GeV, and to lie within $\abs{\eta} < 2.5$, but outside of the transition region between ECAL barrel and endcaps ($1.44 < \abs{\eta} < 1.57$) to avoid low-quality reconstruction due to a gap between the barrel and endcap calorimeters, which is filled with services and cables.
Electron candidates must pass a number of identification and isolation requirements optimized for high-\pt electrons~\cite{Khachatryan:2014fba,Khachatryan:2015hwa,Khachatryan:2016zqb}. These criteria include requirements on the geometrical matching between the ECAL deposit and the reconstructed track, the ratio of energies deposited in HCAL and ECAL calorimeters, the shape of the ECAL deposit, the impact parameters of the track, and the number of missing hits in the silicon tracker.
A requirement on the electron isolation is also applied, which considers tracks originating from the same vertex as the electron, within $\Delta R(\text{electron}, \text{track}) = \sqrt{\smash[b]{(\Delta \eta)^{2} + (\Delta \phi)^{2}}} < 0.3$ of the electron, where $\Delta\eta$ and $\Delta\phi$ are the separations in pseudorapidity and azimuthal angle (in radians), respectively, between the electron and a track. The scalar sum of the \pt of these tracks is required to be less than 5\GeV.

Muons are reconstructed combining tracks in the CMS muon system and inner tracker~\cite{Chatrchyan:2012xi,Sirunyan:2018}.
They are required to have $\pt > 53\GeV$ and $\abs{\eta} < 2.4$.
Muons must satisfy various reconstruction and identification requirements on the impact parameters of the track, the number of hits in the pixel tracker, the number of tracker layers with hits, the relative \pt uncertainty, the number of muon chambers included in the muon track fit, and the number of segments reconstructed in the muon detector planes.
Muons are considered isolated if the scalar sum of the \pt of tracks from the primary vertex within $\Delta R < 0.3$ of the muon is less than 10\% of the \pt of the muon.

Events are required to contain a single lepton (electron or muon). To reject backgrounds from Drell--Yan and fully leptonic \ttbar events, we reject events that contain additional leptons, where the \pt threshold for the additional leptons is lowered to 35 (20)\GeV for the electron (muon) channel, respectively.

Jets are reconstructed from PF particles, clustered by the anti-\kt algorithm~\cite{Cacciari:2008gp,Cacciari:2011ma} with distance parameters of 0.4 and 0.8, denoted as AK4 and AK8 jets, respectively. The momentum of a jet is determined as the vectorial sum of all particle momenta in the jet, and is found from simulation to be, on average, within 5 to 10\% of the true momentum of the generated particles in the jet over the whole \pt distribution and detector acceptance. Additional $\pp$ interactions within the same or nearby bunch crossings can contribute additional tracks and calorimetric energy depositions, increasing the apparent jet momentum. To mitigate this effect, tracks identified to be originating from pileup vertices are discarded prior to the clustering, and an offset correction is applied to correct for remaining contributions~\cite{Khachatryan:2016kdb}. Jet energy corrections are derived from simulation so that the average measured response of jets becomes identical to that of particle level jets. In situ measurements of the energy balance in dijet, photon+jet, $\PZ$+jet, and multijet events are used to determine any residual differences between the jet energy scale in data and in simulation, and appropriate corrections are made~\cite{Khachatryan:2016kdb}.
Additional selection criteria are applied to each jet to remove jets potentially dominated by instrumental effects or reconstruction failures.

The missing transverse momentum vector \ptvecmiss is computed as the negative vector sum of the transverse momenta of all the PF candidates in an event, and its magnitude is denoted as \ptmiss~\cite{CMS-PAS-JME-17-001}. The \ptvecmiss is modified to account for corrections to the energy scale of the reconstructed jets in the event.
The \ptmiss is required to be larger than 110 (40)\GeV in the electron (muon) channel to reject QCD multijet background events.
The higher \ptmiss threshold for the electron channel is necessary to reduce the contribution of QCD multijet events with mismeasured \ptmiss from jets misidentified as electrons, since the electron identification criteria are optimized for greater efficiency at the expense of lower purity.

The leptonic $\PW$ boson candidate is constructed from the lepton and the \ptvecmiss.
The longitudinal momentum of the neutrino can be reconstructed from the $\PW$ boson mass constraint, assuming that the neutrino is the sole contributor to the \ptmiss.
The $x$ and $y$ components of the neutrino momentum therefore come directly from the \ptvecmiss.
Fixing the mass of the $\PW$ boson candidate to its pole mass value, one can relate the four-momentum of the $\PW$ boson to those of the lepton and neutrino via a quadratic equation, which can have two real or complex solutions.
In the case of two real solutions, the solution with the smaller absolute value is assigned as the neutrino longitudinal momentum, whereas in the case of two complex solutions, the real part common to both is instead assigned.
In simulated SM diboson samples, this method assigns the correct solution in approximately 90\% of events.
Although $\PW \to \PGt\PGnGt \to \Pe\PGne/\PGm\PGnGm+\PGnGt$ decays are included in the simulated signals, they are not efficiently reconstructed because of the presence of the second neutrino.
The reconstructed leptonic $\PW$ boson candidate is then required to have $\pt > 200\GeV$.

The AK8 jets with $\pt>200\GeV$ and $\abs{\eta}<2.4$ are used as the basis for the identification of hadronic boson decays, whereas AK4 jets with $\pt>30\GeV$ and $\abs{\eta}<2.4$ are used for the rejection of background processes containing a top quark decay. All jets have to pass basic quality criteria based on the relative fractions of different PF particle types within the jets. They are also excluded from the analysis if they are within $\Delta R < 0.3$ of the lepton.

The AK8 jet with the highest \pt serves as the hadronic $\PW$ or $\PZ$ boson (hereafter denoted by $\PV$) candidate. The leptonic and hadronic boson candidates are combined into a diboson system by adding their four-momenta. The invariant mass of the reconstructed diboson system, $\mwv$, is the chosen event variable for the signal extraction. Because signal events are expected to have a back-to-back topology in the detector, we require events in the signal region to satisfy the following requirements: $\Delta R (\text{AK8 jet}, \text{lepton}) > \pi/2$, $\Delta \phi (\text{AK8 jet}, \ptvecmiss) > 2$, and $\Delta \phi (\text{AK8 jet}, \PW) > 2$, where $\PW$ denotes the reconstructed leptonic $\PW$ boson candidate. Additionally, we require $\mwv > 900\GeV$ to restrict the phase space to a region where the background can be described by a monotonically falling parametric function.

Jets originating from the decay of $\cPqb$ quarks are identified using the combined secondary vertex discriminator~\cite{CMS-BTV-16-002}.
Those AK4 jets fulfilling the tight working point of the discriminator (${>}0.9535$) are considered as $\cPqb$ tagged.
This working point has an overall efficiency of 41\% for correctly identifying a jet from a bottom quark, with a 0.1\% probability of misidentifying a jet from a light-flavour quark or gluon as $\cPqb$ tagged.
Events that contain one or more {\cPqb}-tagged AK4 jets are rejected to reduce the background from processes containing a top quark decay, especially \ttbar.
However, only AK4 jets with a separation of $\Delta R > 0.8$ with respect to the hadronic $\PV$ are included to avoid rejecting $\WZ$ signal events with a $\PZ \to \bbbar$ decay.

To discriminate between AK8 jets originating from heavy-boson decays and jets originating from the hadronization of quarks and gluons, and to improve the resolution of the $\PV$ jet mass and reduce the residual effect of pileup, we employ a suitable jet grooming algorithm~\cite{Chatrchyan:2013vbb,CMS-PAS-JME-14-002}.
In this search, we apply a modified mass-drop algorithm~\cite{jetmass_nnll1,Butterworth:2008iy}, known as the \emph{soft drop} algorithm~\cite{Larkoski:2014wba}, to the AK8 jet, with parameters $\beta=0$, $z_\text{cut}=0.1$, and $R_0 = 0.8$.
This removes soft, wide-angle radiation from the jet, reducing the mass of jets initiated by gluons or single quarks, and improving the jet mass resolution for jets originating from heavy particles, here the $\PW$ and $\PZ$ bosons.
To further improve the jet mass resolution, prior to grooming the pileup per particle identification (PUPPI) algorithm~\cite{Bertolini:2014bba} is used to mitigate the effect of pileup at the reconstructed particle level, making use of local shape information, event pileup properties, and tracking information.
Charged particles identified as originating from pileup vertices are discarded.
For each neutral particle, a local shape variable is computed using the surrounding charged particles that are compatible with the primary vertex and within the tracker acceptance ($\abs{\eta} < 2.5$), and using both charged and neutral particles in the region outside of the tracker coverage.
The momenta of the neutral particles are then scaled by the probability that they originate from the primary interaction vertex deduced from the local shape variable, superseding the need for jet-based pileup corrections~\cite{CMS-PAS-JME-16-003}.
The invariant mass of the resulting jet is the PUPPI soft drop mass \msd, one of the most important variables in this analysis.

To further discriminate against jets from the hadronization of gluons and single quarks, the $N$-subjettiness~\cite{Nsubjet} variable $\taun$ is used.
The $N$-subjettiness variable quantifies the compatibility of clustering the jet constituents into exactly $N$ subjets, with small values representing configurations more compatible with the $N$-subjet hypothesis. The ratio between 2- and 1-subjettiness, $\tau_{21} = \tau_{2}/\tau_{1}$, is a powerful discriminant between jets originating from hadronic {\PV} decays and those from single gluon or quark hadronization.

We require the AK8 jet in the signal region to have $65 < \msd < 105\GeV$ and $\tau_{21} < 0.55$ to suppress the background processes, especially those from \Wjets events.
To better distinguish the $\WW$ and $\WZ$ final states, the signal region is subdivided into the $\WW$-sensitive region ($65 < \msd < 85\GeV$) and the $\WZ$-sensitive region ($85 < \msd < 105\GeV$).
In addition to the signal region, as defined by the selection described above, we define several control regions, each of which is designed to enhance a specific background contribution:
\begin{itemize}
\item \Wjets control region: also referred to as the sideband, defined analogously to the signal region but with $\msd \in [40,65] \cup [105,150]\GeV$. The two intervals define the lower and upper sidebands, respectively.
\item \ttbar control region: defined like the signal region, but requiring at least one {\cPqb}-tagged AK4 jet, and $\msd \in [40,150]\GeV$.
\end{itemize}

The analysis proceeds simultaneously in the electron and muon channels to take into account slight differences in efficiency, acceptance, and background composition.
In each of the two channels, the signal is extracted by a two-dimensional fit to the \msd and \mwv distributions in data, with each signal and background contribution represented by a parametric function.
Minor background contributions are modelled by directly fitting parametric functions to the simulated samples and keeping them fixed in the final fit.
In contrast, major backgrounds are modelled by first determining the function parameters by fitting to the simulation, then using the fit result uncertainties as priors when fitting these to data in the process of the signal extraction.
The fit range in \msd includes the signal region as well as the \Wjets control region, to help constrain the \Wjets background.
To accurately estimate this dominant background, the ratio of the \Wjets \mwv distributions in the signal and \Wjets control regions in data is constrained to match that predicted by the simulation.

\section{Signal modelling}
\label{sec:signalModel}

For diboson processes, with or without additional contributions from anomalous couplings, the $\mwv$ distribution can be modelled to a good approximation by an exponential decay function.
The inclusion of additional contributions from anomalous couplings leads to an increase of events at higher $\mwv$ values.
Therefore, the signal shape is modelled as a sum of exponential terms, with a combination of terms accounting for pure SM and aTGC contributions, as well as SM--aTGC and aTGC--aTGC interference effects.
The pure aTGC term also includes the error function to ensure its effect is only relevant at larger values of $\mwv$.

The complete signal diboson mass distribution, $F_{\text{signal}}(\mwv)$, is described by:
\begin{linenomath}
\begin{equation}
\label{eqSignalModel}
\begin{aligned}
F_{\text{signal}}(\mwv) = & N_{\text{SM}} \left( \re^{a_0\mwv}+\re^{a_{\text{corr}}\mwv} \right) \\
             & + \sum_{i} {\left( N_{\ci,1} \ci^2 \re^{a_{i,1}\mwv} \left( \frac{1+\erf[(\mwv-a_{0,i})/a_{\mathrm{w},i}]}{2} \right)  + N_{\ci,2} \ci \re^{a_{i,2}\mwv} \right) } \\
             & + \sum_{i< j} { \left( N_{\ci,\cj} \ci \cj \re^{a_{{ij}}\mwv} \right) },
\end{aligned}
\end{equation}
\end{linenomath}
where $\ci$ are the various aTGC parameters, and $\erf$ is the error function.
The complete signal distribution can be decomposed into four contributions: the SM part with no dependence on $\ci$, pure aTGC contributions proportional to $\ci^2$, aTGC--SM interference terms proportional to $\ci$, and bilinear interference terms between the different aTGCs proportional to $\ci \cj$ for $i \neq j$.
The parameters $N_{\text{SM}}$, $N_{\ci,1}$, $N_{\ci,2}$, and $N_{\ci,\cj}$ are the normalization of the SM, pure aTGC, aTGC--SM interference, and aTGC--aTGC interference terms for the various $c_{i,{j}}$, respectively.
Similarly, $a_0$, $a_{i,1}$, $a_{i,2}$, and $a_{{ij}}$ are the exponential decay constants of each of these contributions.
The parameters $a_{0,i}$ and $a_{\mathrm{w},i}$ govern the turn-on position and steepness of the error function in the pure aTGC contribution for a given $\ci$.
The exponential term with decay constant $a_{\text{corr}}$ is a small correction added to account for the deviation of the SM contribution from a simple exponential at higher values of $\mwv$.

These parameters are determined empirically from the signal simulation.
This is done to facilitate easier interpolation between aTGC parameter values, and to avoid large statistical uncertainties from regions with limited numbers of MC events.
The following procedure is used to extract the various slope and normalization parameters.
First, the SM shape and normalization parameters $a_0$ and $N_{\text{SM}}$ are extracted from the simulation by reweighting the \MGvATNLO signal simulation (which is generated with aTGCs) to the SM simulation (without any aTGCs).
Then, the aTGC--SM interference parameters $a_{i,2}$ and $N_{\ci,2}$ are derived by comparing the shapes when an aTGC is set to equal values but with opposite signs.
The pure aTGC parameters $a_{i,1}$, $a_{0,i}$, $a_{\mathrm{w},i}$, and $N_{\ci,1}$ are then extracted in a simultaneous fit of the SM, aTGC--SM interference, and pure aTGC terms to samples weighted with only a single, nonzero aTGC.
Finally, the aTGC--aTGC interference terms are derived by comparing samples with pairs of aTGCs set to nonzero values.
The error function in the pure aTGC terms is introduced to accurately model the turn-on behaviour of the aTGC contributions.
To simplify the signal model, very small contributions from $\cwww$--SM interference, $\cwww$--$\cb$ interference, and the error function for $\cb$ in the $\WZ$ region are neglected.

\section{Background modelling}
\label{sec:backgroundModel}

There are two major contributions to the SM background (\Wjets, and \ttbar), and two minor contributions (single top quark and SM diboson production).
Even with substantial enhancements of the diboson cross section in the event of nonzero aTGCs, any signal contribution in the control regions is expected to be small since the control regions are explicitly designed to enrich the backgrounds whilst rejecting contamination from the signal processes.

The normalizations of the background contributions are determined during the signal extraction through a two-dimensional fit to the $(\msd, \mwv)$ distributions in data.
The \msd and \mwv shape parameters of the \Wjets background, along with the \mwv shape parameters of the \ttbar background, are also extracted from the two-dimensional fit, as described below.
The \msd shape of the \ttbar background, as well as the shapes of the single top quark and SM diboson background contributions, are taken directly from fits to simulation.

\begin{figure}[htb]
\centering
\includegraphics[width=0.49\textwidth]{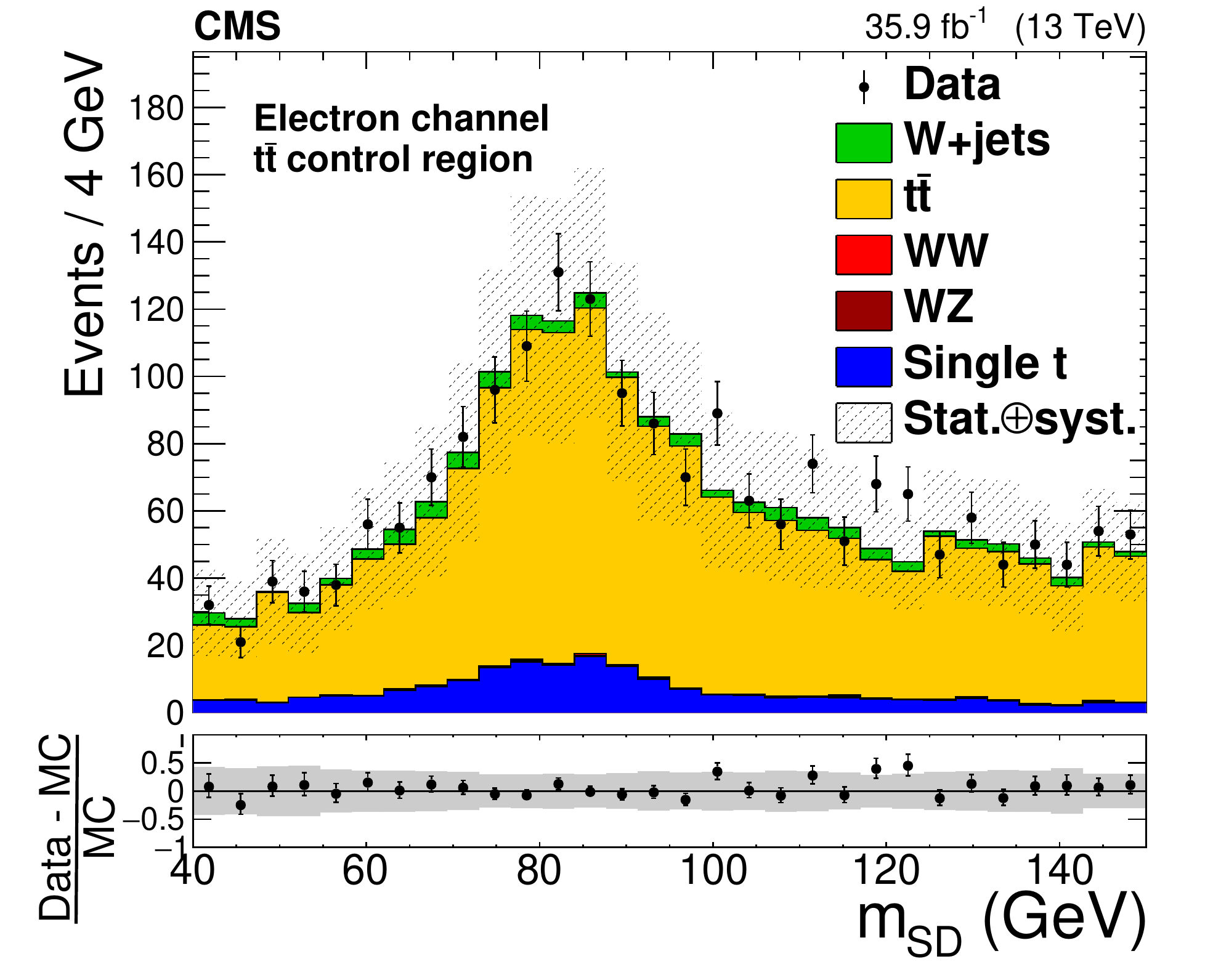}
\includegraphics[width=0.49\textwidth]{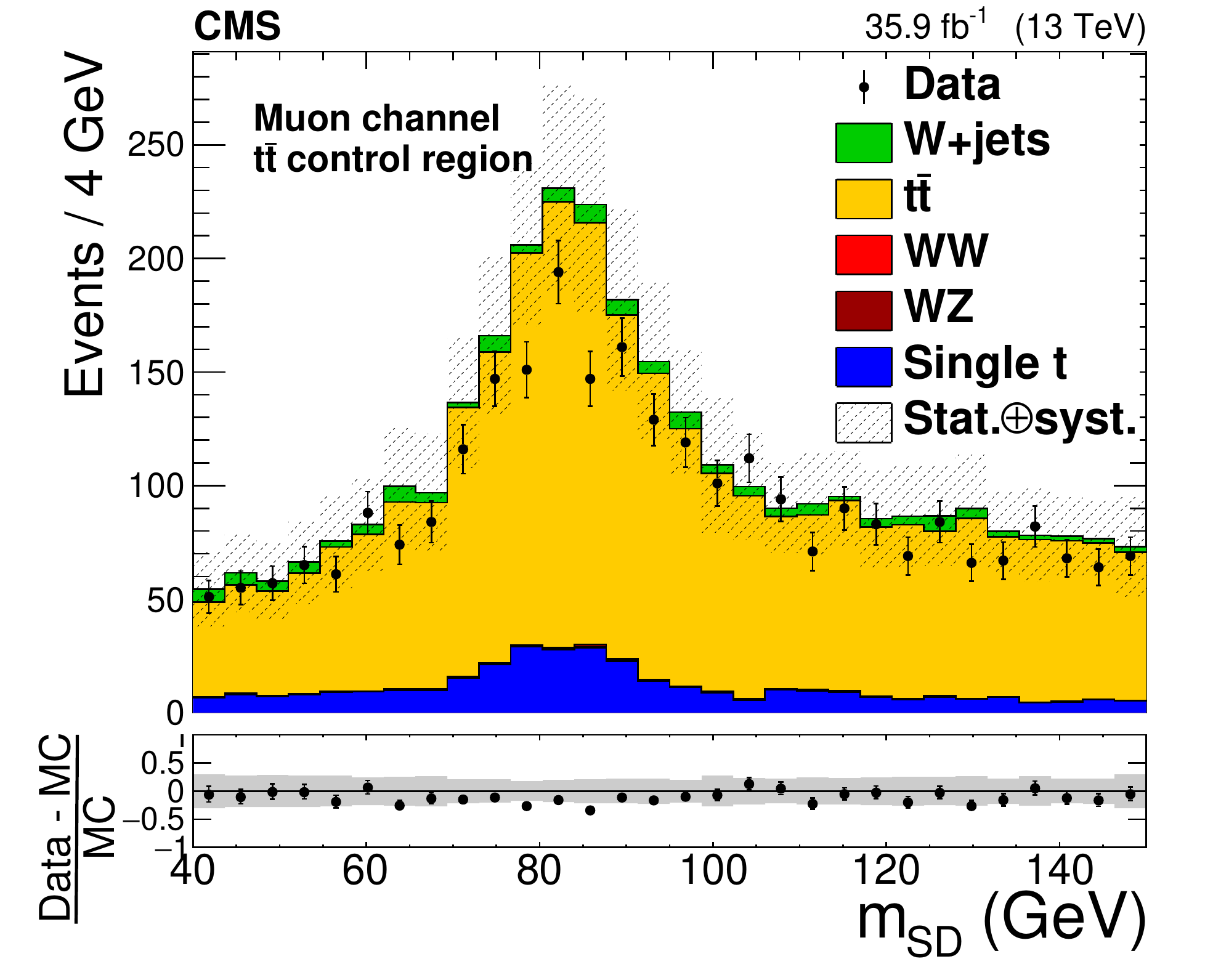} \\
\includegraphics[width=0.49\textwidth]{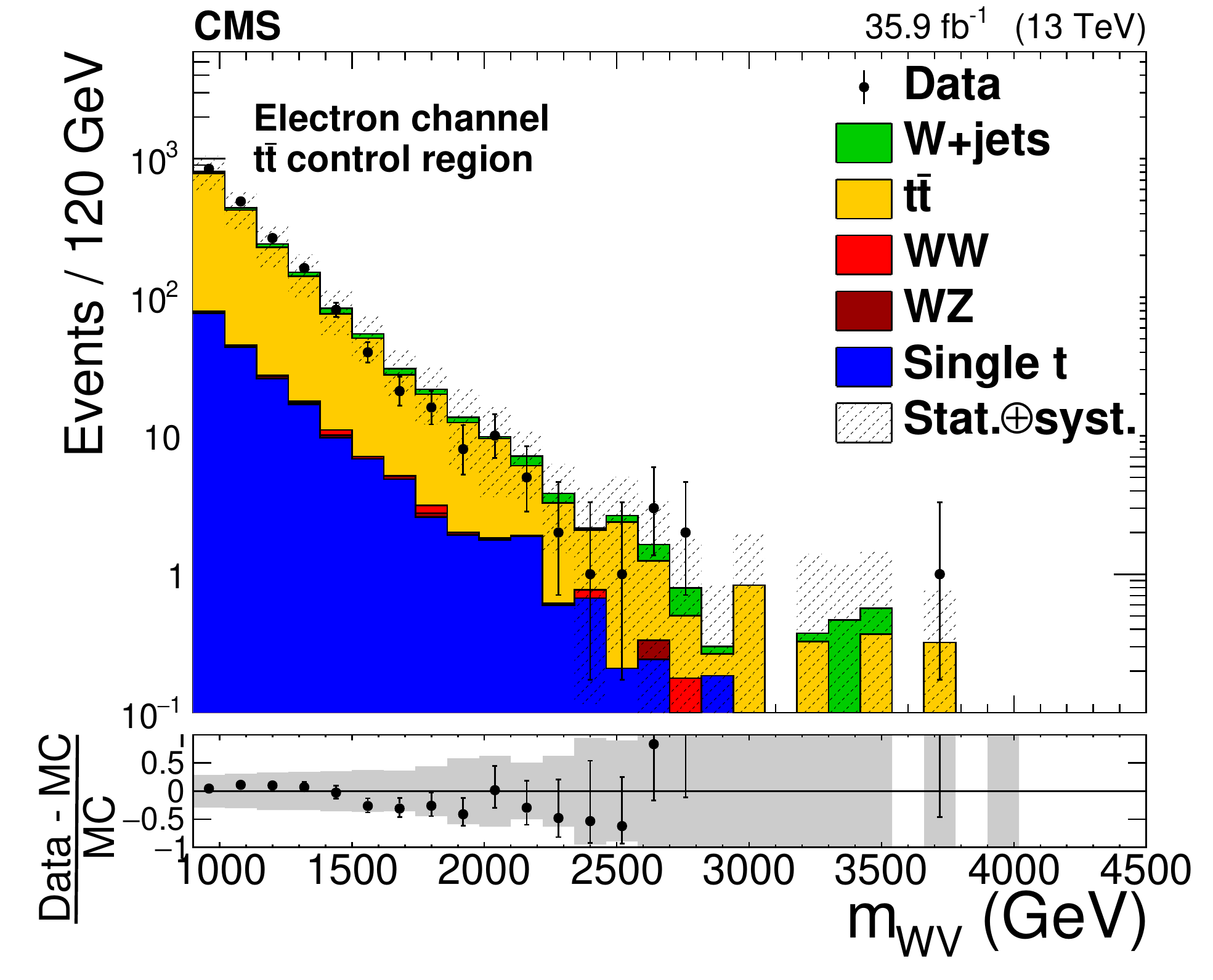}
\includegraphics[width=0.49\textwidth]{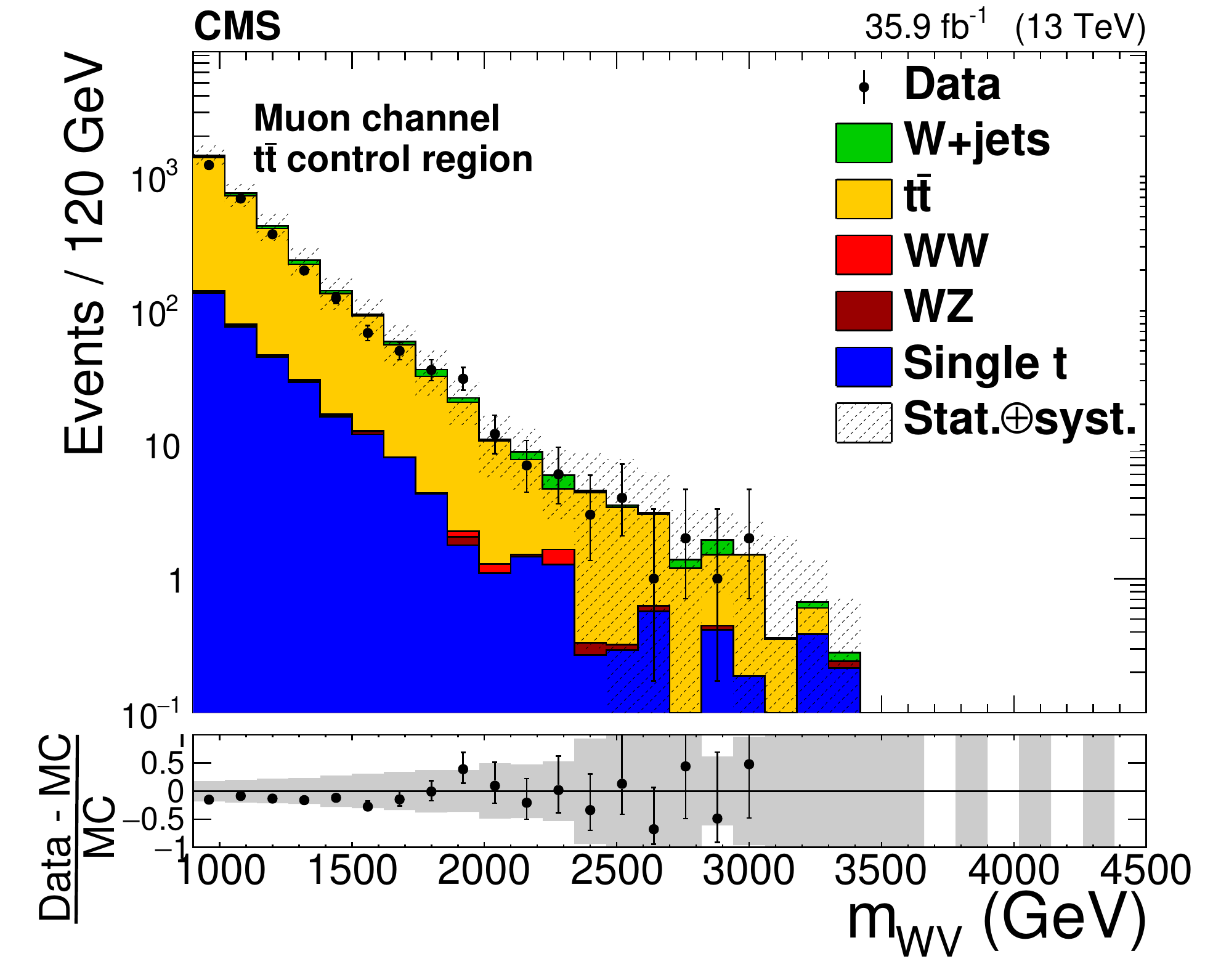}
\caption{Comparison between data and simulation for the \msd (upper) and \mwv (lower) distributions in the \ttbar control region. Contributions from simulation are normalized to the total integrated luminosity of the data using their respective SM cross sections. The electron channel is shown on the left, while the muon channel is shown on the right. The lower panel in each figure shows the relative difference between data and simulation.
The light grey hashed region in the main panels and dark grey band in the lower ratio panels represent the combined statistical and systematic uncertainties, with details of the latter discussed in Section~\ref{sec:syst}.}
\label{fig:ttbarControl}
\end{figure}

Since the \ttbar background estimate is largely based on a template derived from simulation, we validate its accuracy by verifying that data in the \ttbar control region are well-modelled by the simulation.
Of particular importance are the \msd and \mwv distributions, since \mwv is used to extract limits on anomalous couplings.
Figure~\ref{fig:ttbarControl} shows that the simulation is in agreement with the data for these variables in the \ttbar control region, which is verified by a $\chi^2$ test (with $p$-value ${>}0.99$ in all cases).

Because of the lack of knowledge of the continuous dependence of the shape parameters describing the $\mwv$ distribution as a function of $\msd$, and with no reliable way to continuously model it, the two-dimensional fit is constructed by defining four separate regions in \msd (lower sideband, signal $\WW$, signal $\WZ$, and upper sideband).
All four regions are fitted simultaneously to constrain the shape parameters.
In each region, the shape parameters describing the $\mwv$ distribution are constant with respect to $\msd$.
In the sideband regions these shape parameters are determined by fitting to the data.
In the signal regions the shape parameters are instead obtained by assuming that the simulation accurately describes the ratio of the $\mwv$ distributions in the signal and sideband regions. This ratio function $\alpha^{\MC}(\mwv)$ is used to transfer the shape of the \Wjets background, which comes from data, from the sideband to the signal regions, thereby encoding the dependence of $\mwv$ on $\msd$ (the $\alpha$ ratio method~\cite{Khachatryan:2014gha,Sirunyan:2016cao}).
The total background contribution in the signal region, $F_{\text{bkg}}^{\SR}$, can therefore be expressed as:
\begin{linenomath}
\begin{equation}
\begin{aligned}
F_{\text{bkg}}^{\SR}(\mwv) &= F_{\Wjets}^{\SB,\ \text{data}} \alpha^{\MC}(\mwv) + F_{\ttbar}^{\SR,\ \MC} + F_{\text{single $\cPqt$}}^{\SR,\ \MC} + F_{\text{diboson}}^{\SR,\ \MC}\\
        \alpha^{\MC}(\mwv) &= \frac{F_{\Wjets}^{\SR,\ \MC}}{F_{\Wjets}^{\SB,\ \MC}},
\end{aligned}
\end{equation}
\end{linenomath}
where $F$ denotes the parametric functions representing various background contributions in the signal (SR) and sideband (SB) regions.
The statistical uncertainties from the fits to data and simulation are propagated to the final prediction of the \Wjets and \ttbar backgrounds, as discussed in Section~\ref{sec:syst}.

In the fit, the various background contributions have different constraints placed upon their normalizations and $\mwv$ and $\msd$ shape parameters, depending on the importance of the contribution, and the level of certainty in its modelling.
The normalization and \msd shape parameters of the \Wjets contribution are allowed to vary without constraint to account for possible mismodelling.
The \mwv shape parameters for this contribution are allowed to vary within their uncertainties, which arise from the uncertainties in the simulation entering $\alpha^{\MC}$.
The normalization and $\mwv$ shape parameters of the \ttbar contribution are also allowed to vary within their respective uncertainties; however the $\msd$ shape parameters are kept fixed.
For the single top quark contribution, the shape parameters for both $\mwv$ and $\msd$ are kept fixed, whilst the normalization is allowed to float within the systematic uncertainty.
Similarly, the shape parameters of the SM diboson contribution are also kept fixed.
However, the normalization is constrained with 100\% uncertainty to cover its systematic uncertainty, and to allow for a substantial contribution from aTGC processes, consistent with the sensitivity of this analysis.
Further discussion of the sources contributing to these uncertainties is provided in Section~\ref{sec:syst}.
Normalization values before and after the fit for all contributions are shown in Table~\ref{tab:fitnormtab}.

\begin{table}[!htb]
        \centering
        \topcaption{Results of the signal extraction fits. The uncertainties in the pre-fit yields are their respective pre-fit constraints, whilst the uncertainties in the post-fit yields are the corresponding total post-fit uncertainties. Since the normalization of the \Wjets contribution is allowed to vary freely in the fit, it does not have any corresponding pre-fit uncertainties.}
        \cmsTable{
        \begin{tabular}{lcccccc}
                & \multicolumn{3}{c}{Electron channel} & \multicolumn{3}{c}{Muon channel} \\
                & Pre-fit & Post-fit & Scale factor & Pre-fit & Post-fit & Scale factor \\
                \hline
                \Wjets     & 2421    & $3036 \pm 123$ & 1.25 & 4319 &   $4667 \pm 182$  & 1.08 \\
                \ttbar     & $1491 \pm 324$ & $1127 \pm 119$ & 0.76 & $2632 \pm 570$ & $1978 \pm 202$  & 0.75 \\
                Single $\cPqt$ & $271 \pm 39$  & $242 \pm 26$  & 0.89 & $509 \pm 69$  & $449 \pm 43$   & 0.88 \\
                Diboson    & $314 \pm 314$ & $267 \pm 102$ & 0.85 & $552 \pm 552$ & $465 \pm 162$  & 0.84 \\ [\cmsTabSkip]
                Total expected    & 4497  &  $4672 \pm 201$ & 1.04 & 8012 & $7559 \pm 319$ & 0.94 \\[\cmsTabSkip]
                Data       & \multicolumn{3}{c}{4691} & \multicolumn{3}{c}{7568}  \\
                \hline
        \end{tabular}
        }
        \label{tab:fitnormtab}
\end{table}

\section{Systematic uncertainties}
\label{sec:syst}

There are several systematic uncertainties that affect the normalizations of the \ttbar, single top quark, and diboson processes that are derived from simulation. These uncertainties are included in the final fit to the data.

An uncertainty of 2.5\%~\cite{lumi} is included to account for the uncertainty in the integrated luminosity measurement of the 2016 data set. This uncertainty is treated as correlated between the different processes.

The uncertainty associated with the pileup reweighting of simulated events is calculated from the uncertainty in the total inelastic cross section that is used to derive the pileup weights~\cite{Sirunyan:2018nqx}.

We include uncertainties in the cross section calculations used to normalize the contributions from simulation. This is done by utilizing the uncertainties associated with the PDFs following the recommendations of the PDF4LHC working group~\cite{Butterworth:2015oua}. Uncertainties corresponding to the choice of renormalization and factorization scales (\muR and \muF, respectively) are computed by reweighting the simulated samples for all combinations of nominal scales and scales multiplied/divided by a factor of two, excluding combinations in which one scale is increased and the other simultaneously decreased, and using the largest deviation as the uncertainty.

A normalization uncertainty of 14\% describing the mismodelling of the $\tau_{21}$ selection efficiency~\cite{Sirunyan:2016cao} is applied to all contributions derived from simulation containing hadronic $\PV$ boson decays, and is treated as correlated between the different processes. This uncertainty is not applied to the \Wjets contribution, which is directly estimated from data, nor to the $t$- and $s$-channel subprocesses of single top quark production, where the hadronically decaying $\PV$ boson candidate is associated with jets arising from the hadronization of a single light quark or gluon.

For the \ttbar and $\WZ$ samples, we include the uncertainties in the efficiencies to identify and misidentify (mistag) {\cPqb} quark jets~\cite{CMS-BTV-16-002}. The uncertainties in the {\cPqb} tagging efficiencies most notably affect the normalization of the \ttbar background, whereas the misidentification uncertainties have only a small impact across the samples.

Uncertainties in the jet energy scale have been measured~\cite{Khachatryan:2016kdb}, and are propagated by varying the jet energy scale within its uncertainty for both AK4 and AK8 jets, simultaneously.
Similarly, uncertainties in the jet energy resolution are applied to both AK4 and AK8 jets simultaneously by varying their resolutions by ${\pm}1$ standard deviation.

The lepton energy scale is varied within its uncertainty, and its effect is propagated to the signal extraction fit. Lepton resolution uncertainties are included in a similar manner.

Uncertainties in the measurement of lepton efficiency and identification scale factors are also considered. An additional uncertainty is added to account for additional uncertainty in the scale factors at higher electron energies. In the barrel region this uncertainty is 1\% below 90\GeV, 2\% between 90\GeV and 1\TeV, and 3\% above 1\TeV; in the endcaps it is 1\% below 90\GeV, 2\% between 90 and 300\GeV, and 4\% above 300\GeV. In the muon channel, an additional 1\% uncertainty is added related to the muon identification criteria, 0.5\% related to the isolation requirements, and 0.5\% related to the single-muon triggers.

Jet and lepton uncertainties are also propagated to the calculation of \ptmiss.
In addition, the influence of PF candidates not associated to any reconstructed physics object~\cite{Sirunyan:2019kia} (``unclustered'' energy deposits) on \ptmiss are evaluated and propagated as normalization uncertainties.

The normalization uncertainties for the contributions derived from simulation are summarized in Table~\ref{tab:syst}.
The influence of jet and lepton uncertainties on \ptmiss are included in the corresponding jet and lepton uncertainty rows, whilst the \ptmiss uncertainty value is that arising solely from unclustered energy deposits.
\begin{table}[htb]
\centering
\topcaption{Estimated normalization uncertainties (\%) for SM background contributions derived from simulation.}
\begin{tabular}{lcccccccccc}
 & \multicolumn{4}{c}{Electron channel} & & \multicolumn{4}{c}{Muon channel} \\
Uncertainty source & \ttbar & Single $\cPqt$ & $\WW$ & $\WZ$ & & \ttbar & Single \cPqt & $\WW$ & $\WZ$ \\
& \\[\dimexpr-\normalbaselineskip+2pt]
\hline
& \\[\dimexpr-\normalbaselineskip+2pt]
PDF & 2.79  & 0.22 & 1.93 & 2.44 & & 2.71 & 0.25 & 1.78 & 2.54  \\
\muR, \muF & 17.99 & 0.94 & 5.77 & 4.82 & & 17.74 & 1.06 & 5.99 & 4.26  \\
Luminosity & 2.5 & 2.5 & 2.5 & 2.5 & & 2.5 & 2.5 & 2.5 & 2.5   \\
Pileup & 0.59 & 0.29 & 0.90 & 1.40 & & 0.40 & 0.41 & 0.82 & 0.67  \\
$\PV$ tag & 14 & 14 & 14 & 14 & & 14 & 14 & 14 & 14 \\
{\cPqb} tag & 1.05 & 0.85 & 0.04 & 0.08 & & 1.04 & 0.84 & 0.03 & 0.08  \\
{\cPqb} mistag & 0.04 & 0.05 & 0.02 & 0.04 & & 0.05 & 0.05 & 0.03 & 0.04  \\
Jet energy scale & 4.41 & 4.94 & 4.26 & 2.44 & & 3.54 & 2.97 & 3.75 & 2.50  \\
Jet energy resolution & 1.79 & 3.44 & 1.85 & 2.69 & & 0.85 & 0.91 & 0.62 & 2.92  \\
Lepton energy scale & 0.80 & 1.45 & 1.53 & 0.94 & & 0.68 & 1.14 & 1.72 & 1.19  \\
Lepton energy resolution & 0.26 & 1.22 & 0.11 & 0.21 & & 0.02 & 0.27 & 0.14 & 0.33  \\
Lepton ID & 2.12 & 2.22 & 2.30 & 2.26 & & 1.81 & 2.04 & 2.55 & 2.42  \\
\ptmiss & 0.91 & 1.50 & 1.01 & 0.64 & & 0.59 & 0.99 & 0.24 & 0.17  \\[\cmsTabSkip]
Total & 23.74 & 15.84 & 16.44 & 15.91 & & 23.30 & 14.85 & 16.31 & 15.80 \\
\hline
\end{tabular}
\label{tab:syst}
\end{table}

Shape uncertainties for the \Wjets and \ttbar contributions, as well as for the signal model, are also considered.
The shape uncertainty in the \Wjets sideband estimate is propagated from the simultaneous fit of the data sideband, and signal and sideband regions in simulation. The effect of an alternative fit function is also included by inflating the parameter uncertainties to cover the estimate from the alternative function.

The shape uncertainty for the \ttbar contribution is estimated using the uncertainties in the parameters from the fit of the \ttbar shape to simulation. These are included as nuisance parameters in the background model for the final signal extraction.

For the signal modelling, we include statistical uncertainties from the signal modelling procedure described in Section~\ref{sec:signalModel}, as well as shape variations from the PDF, \muR and \muF scales, jet energy scale and resolution, lepton energy scale, \ptmiss, and {\cPqb} tagging-related uncertainties.
Uncertainties in the slope parameters of the exponential functions are derived by extracting the signal model from signal simulation with the relevant conditions varied, and using the difference between the fitted slope parameters for the nominal and varied samples.
The total uncertainty from these shape variations is the sum of all individual uncertainties in quadrature, resulting in a total uncertainty of approximately 5\% for all aTGCs.
This is dominated by the PDF and \muR and \muF scale uncertainties, with smaller contributions from experimental sources, particularly jet energy scale and resolution, and lepton identification, depending on the lepton flavour and signal region under consideration.

Differential corrections from the consideration of higher order NNLO (QCD)~\cite{Grazzini:2016ctr,Grazzini:2017ckn} and NLO (electroweak)~\cite{Kallweit:2017khh} contributions have previously been calculated, and each can be considerable at large \mwv (${\gtrsim}20\%$), larger than the scale uncertainty at NLO (QCD).
However, since the two corrections have opposite signs, they partially cancel out, reducing their overall effect.
In addition, the impact and validity of these higher order corrections on processes with aTGCs has not been fully investigated.
Therefore, they are not considered as an additional source of uncertainty.

\section{Results}
\label{sec:results}

\begin{figure}[h!]
\centering
\includegraphics[width=0.49\textwidth]{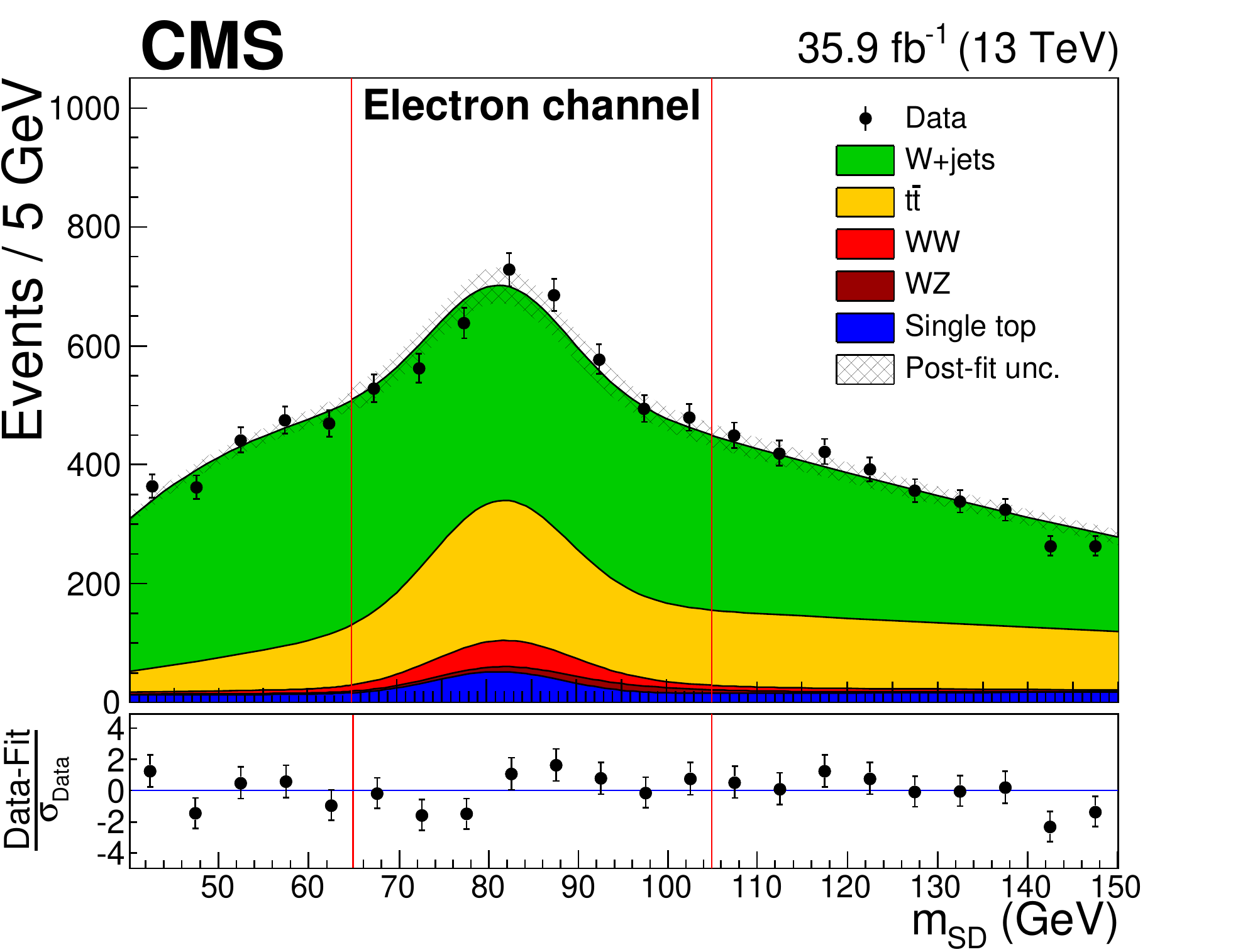}
\includegraphics[width=0.49\textwidth]{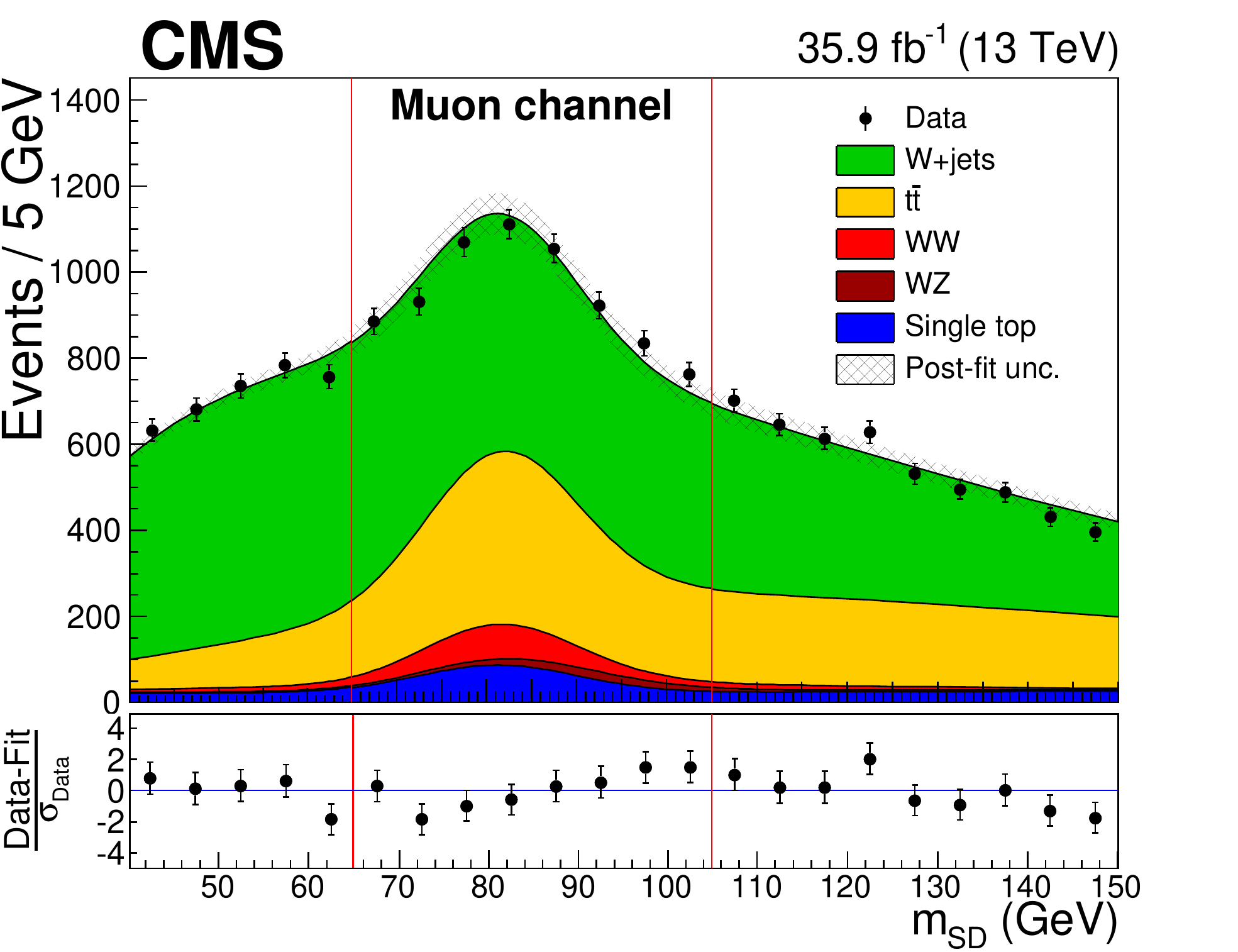}
\caption{Final result of the two-dimensional fit in the electron (left) and muon (right) channels, showing the \msd distribution.}
\label{fig:postfitMSD}
\end{figure}

\begin{figure}[h!]
\centering
\includegraphics[width=0.49\textwidth]{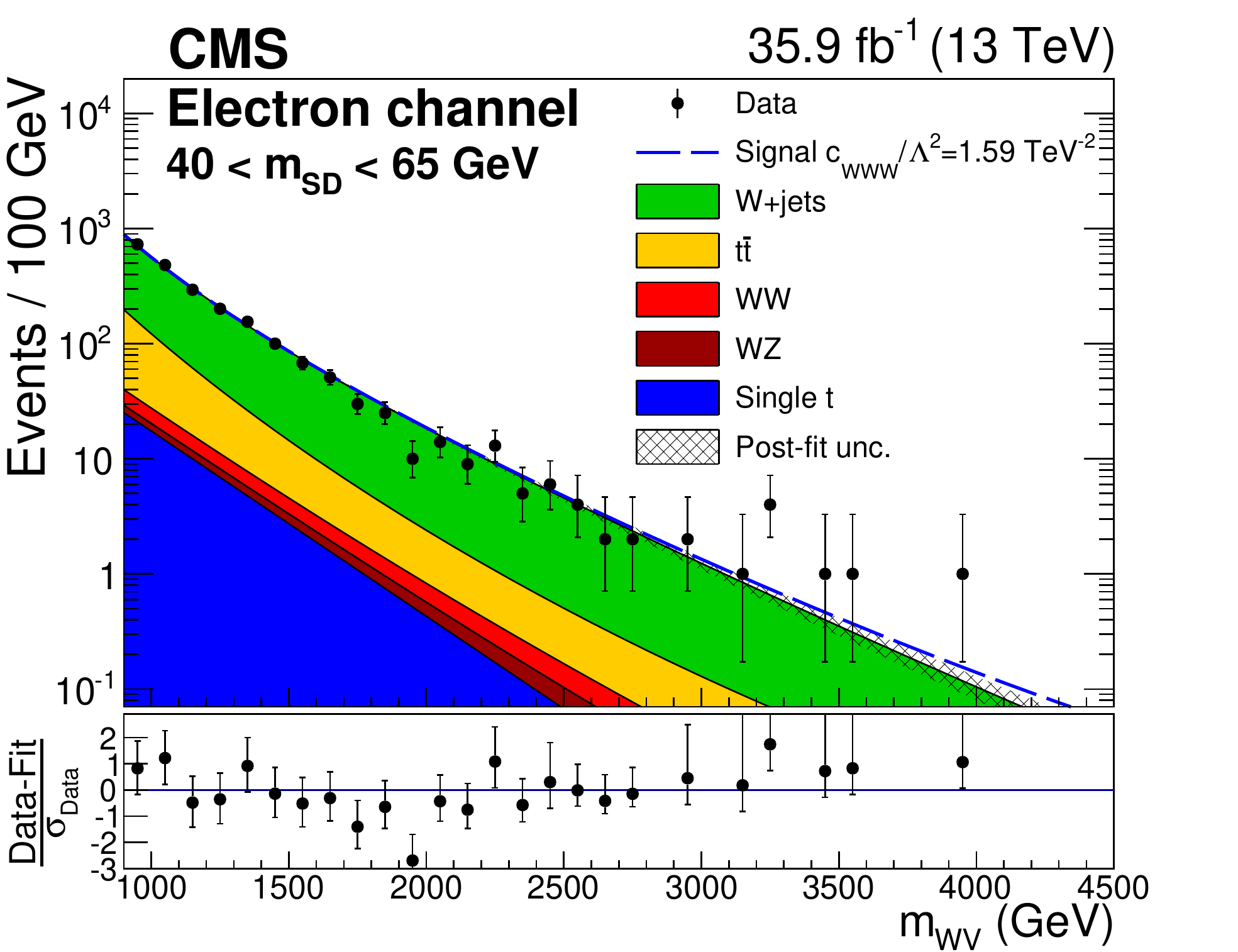}
\includegraphics[width=0.49\textwidth]{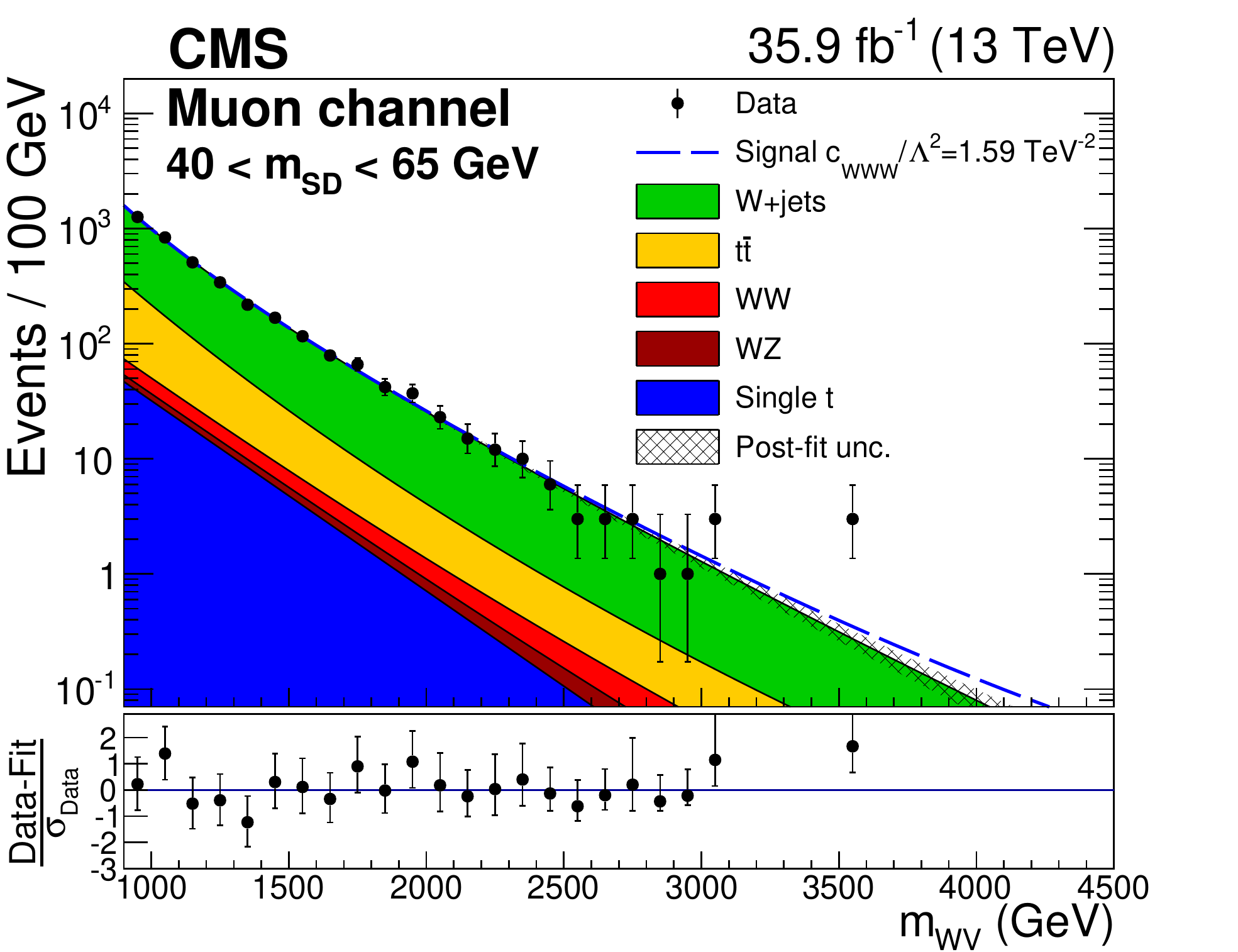} \\
\includegraphics[width=0.49\textwidth]{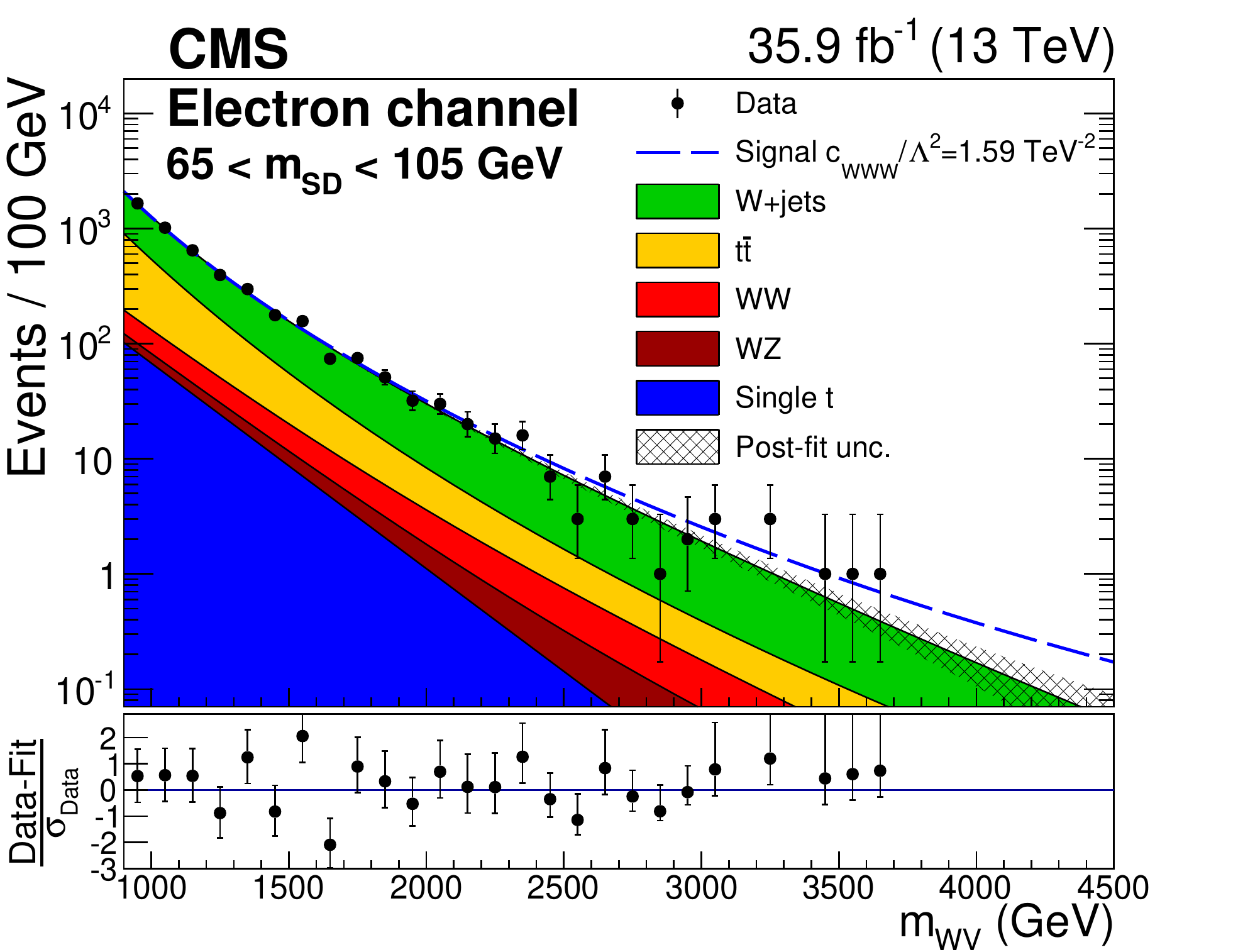}
\includegraphics[width=0.49\textwidth]{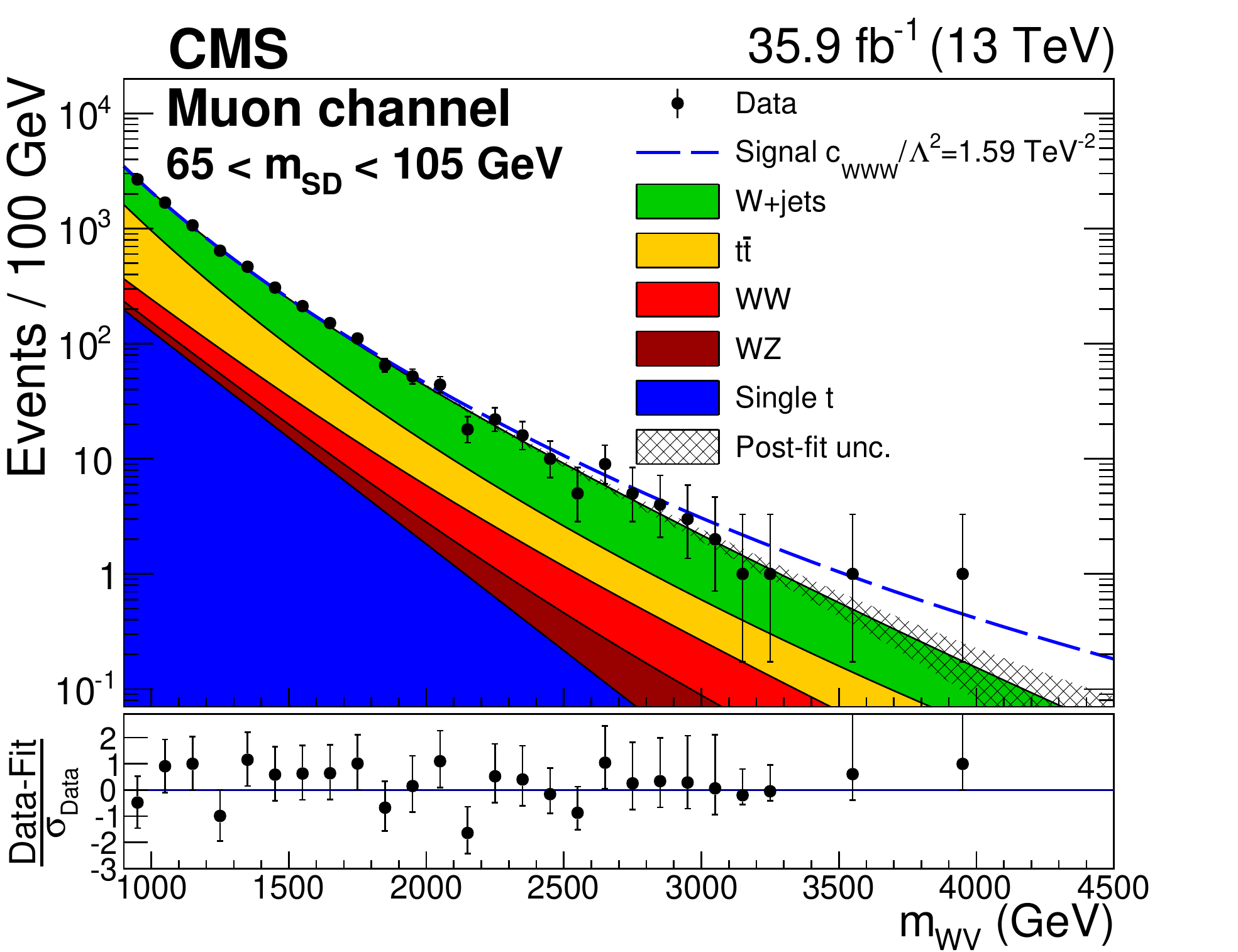} \\
\includegraphics[width=0.49\textwidth]{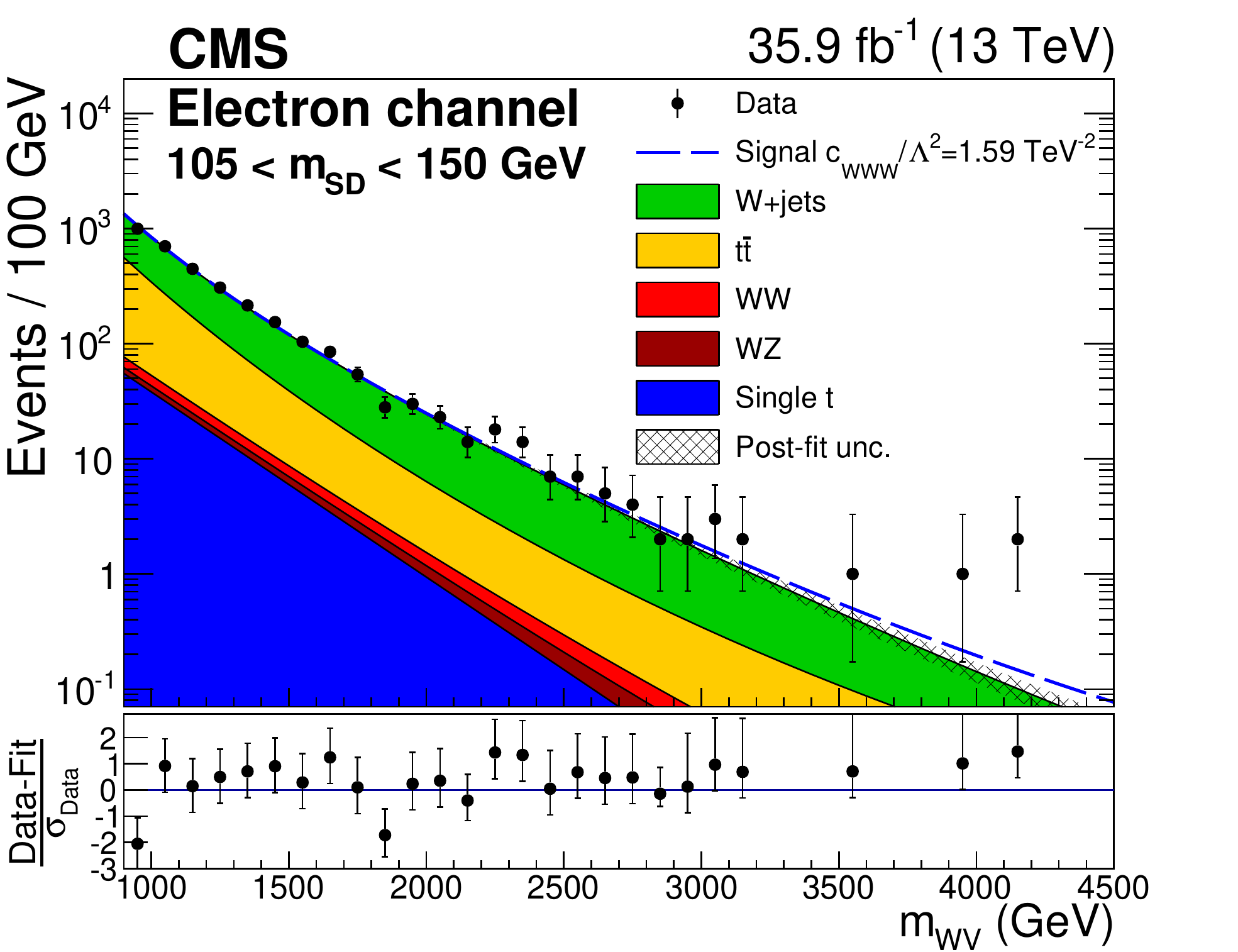}
\includegraphics[width=0.49\textwidth]{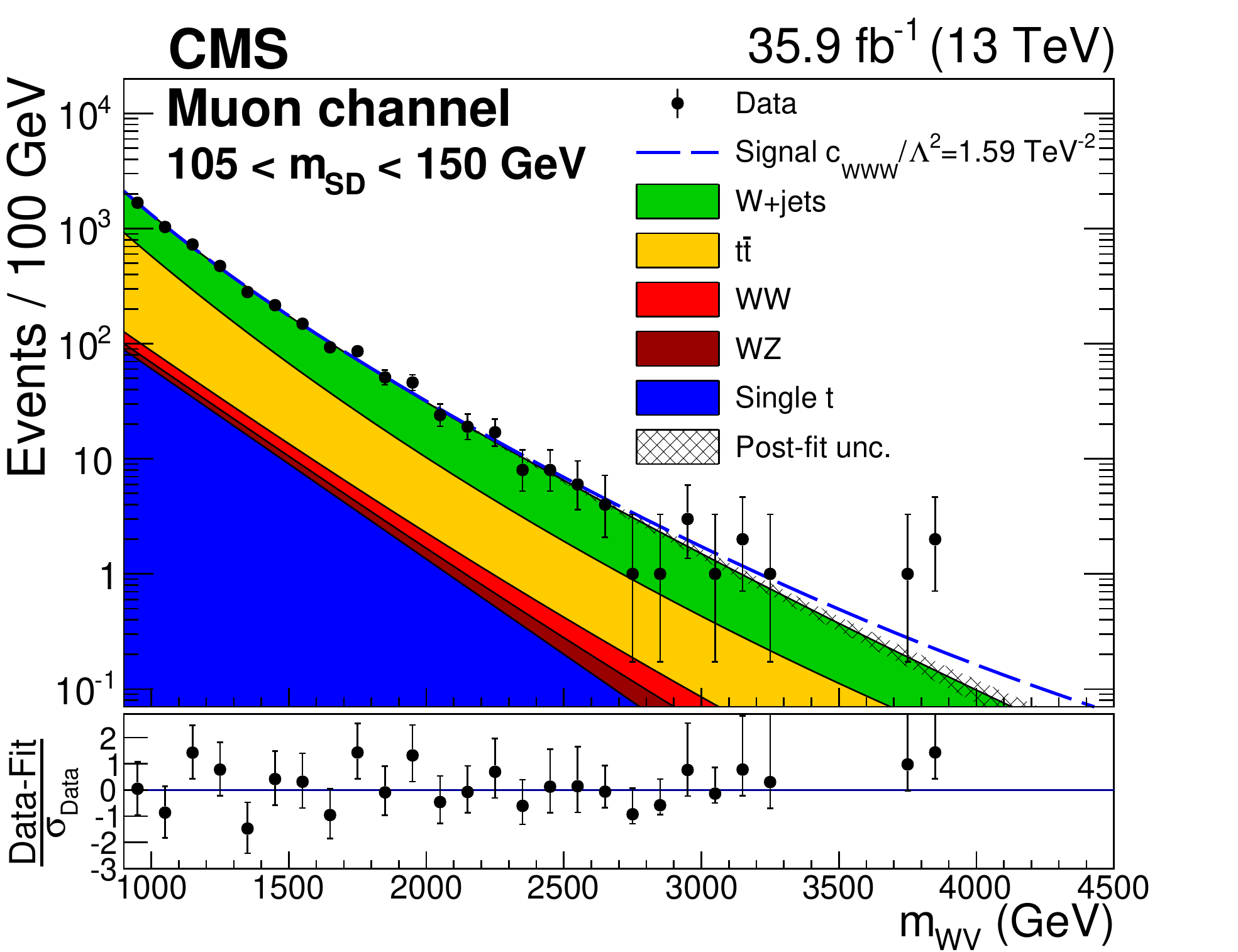} \\
\caption{Final result of the two-dimensional fit in the electron (left) and muon (right) channels, showing the \mwv distributions. The lower sideband, signal, and upper sideband regions are shown on the top, middle, and bottom, respectively. An example of the excluded signal ($\cwww/\Lambda^2 = 1.59\TeV^{-2}$) is represented by the dashed line.}
\label{fig:postfitMWV}
\end{figure}

We set limits on the aTGCs using the data in the signal region and the background estimates.
These are shown in Table~\ref{tab:background:bkgnorms}, and also in Figs.~\ref{fig:postfitMSD} and~\ref{fig:postfitMWV}, where the $\WW$ and $\WZ$ signal regions are combined into one figure for each lepton channel.
Limits are set at 95\% confidence level (\CL) using a simultaneous unbinned maximum likelihood fit of the two-dimensional (\msd, \mwv) distributions in both the electron and muon channels.
The fit to \mwv covers the range $900 < \mwv < 4500\GeV$, where the lower limit is the minimum requirement on \mwv and the upper limit is chosen based on data seen in the control regions.
The best-fit values of the aTGC parameters, along with their confidence intervals, are obtained using scans of the profile likelihood ratio, using the procedure described in Section 3.2 of~\cite{Khachatryan:2014jba}.
Systematic uncertainties are included as nuisance parameters: normalization uncertainties are treated as multiplicative parameters constrained by a log-normal distribution, while shape parameter uncertainties are constrained by Gaussian functions around their nominal values. Limits are derived from the contours of the negative logarithmic likelihood as a function of the aTGCs.

\begin{table}[htb]
        \centering
        \topcaption{Summary of background, signal, and data yields in the $\WW$ and $\WZ$ categories for each lepton channel. Uncertainties in the background contributions are described in Section~\ref{sec:syst}. The diboson signal predictions with anomalous couplings include both standard model and anomalous contributions, as well as the relevant interference terms.}
        \begin{tabular}{lcccc}
                & \multicolumn{2}{c}{Electron channel} & \multicolumn{2}{c}{Muon channel} \\
                & $\WW$ & $\WZ$ &$\WW$ & $\WZ$ \\
                \hline
                \Wjets     & $1618 \pm 66$   & $1418 \pm 57$   & $2529 \pm 99$ & $2138 \pm 83$  \\
                $\ttbar$   & $600 \pm 63$   & $526 \pm 56$   & $1040 \pm 106$   & $938 \pm 96$ \\
                Single top quark & $145 \pm 16$  & $97 \pm 10$  & $264 \pm 25$  & $185 \pm 18$  \\
                Diboson (SM)  & $144 \pm 52$  & $122 \pm 52$ & $265 \pm 88$  & $200 \pm 79$ \\ [\cmsTabSkip]

                Total expected (SM) & $2507 \pm 106$  & $2163 \pm 96$  & $4098 \pm 172$ & $3461 \pm 151$ \\ [\cmsTabSkip]

                Diboson $(\cwww/\Lambda^2 = 3.6\TeV^{-2})$ & $193 \pm 15$ & $185 \pm 15$ & $334 \pm 26$ & $287 \pm 22$ \\
                Diboson $(\cw/\Lambda^2 = 4.5\TeV^{-2})$  & $163 \pm 14$ & $154 \pm 15$ & $283 \pm 23$ & $237 \pm 21$ \\
                Diboson $(\cb/\Lambda^2 = 20\TeV^{-2})$  & $188 \pm 21$ & $144 \pm 14$ & $322 \pm 33$ & $221 \pm 20$ \\ [\cmsTabSkip]

                Data   & 2456 & 2235 & 3996 & 3572 \\
                \hline
        \end{tabular}
        \label{tab:background:bkgnorms}
\end{table}

Limits on individual anomalous couplings are derived by setting the other two couplings to zero.
We assume that the EFT parametrization used here is valid at the energies relevant for this experiment, \ie the true scale associated with any new particles is much larger than the scale $\Lambda$ to which the experiment is sensitive. Specifically, this is possible if the underlying dynamics is strongly coupled~\cite{Contino:2016jqw}.
In addition to the EFT parametrization described in Eq.~(\ref{eq:atgclagrangian}), limits are also computed in terms of the parametrization derived for $\WW$ searches at LEP~\cite{Hagiwara:1986vm,Schael:2013ita} (hereafter referred to as the LEP parametrization), which was also used for searches at the Tevatron~\cite{Abazov:2012ze}. The signal model is reparametrized in terms of the vertex parameters $\lambdaz$, $\deltagz$, and $\deltakapz$ using the relationships in terms of $\cw$, $\cb$, and $\cwww$ given in Ref.~\cite{Degrande:2012wf}, and the likelihood minimization procedure repeated. The resulting limits from both parametrizations, along with the best-fit values, are shown in Table~\ref{tab:1dlim_vertex}.
Limits on the same parameters from $\pp$ collision data taken at a centre-of-mass energy of 8\TeV~\cite{Sirunyan:2017bey} are also quoted to demonstrate the improvement in this analysis (where the limit on $\deltakapg$ has been converted to a limit on $\deltakapz$ using the relationships in Ref.~\cite{Degrande:2012wf}).

\begin{figure}[h!]
\centering
\includegraphics[width=0.32\textwidth]{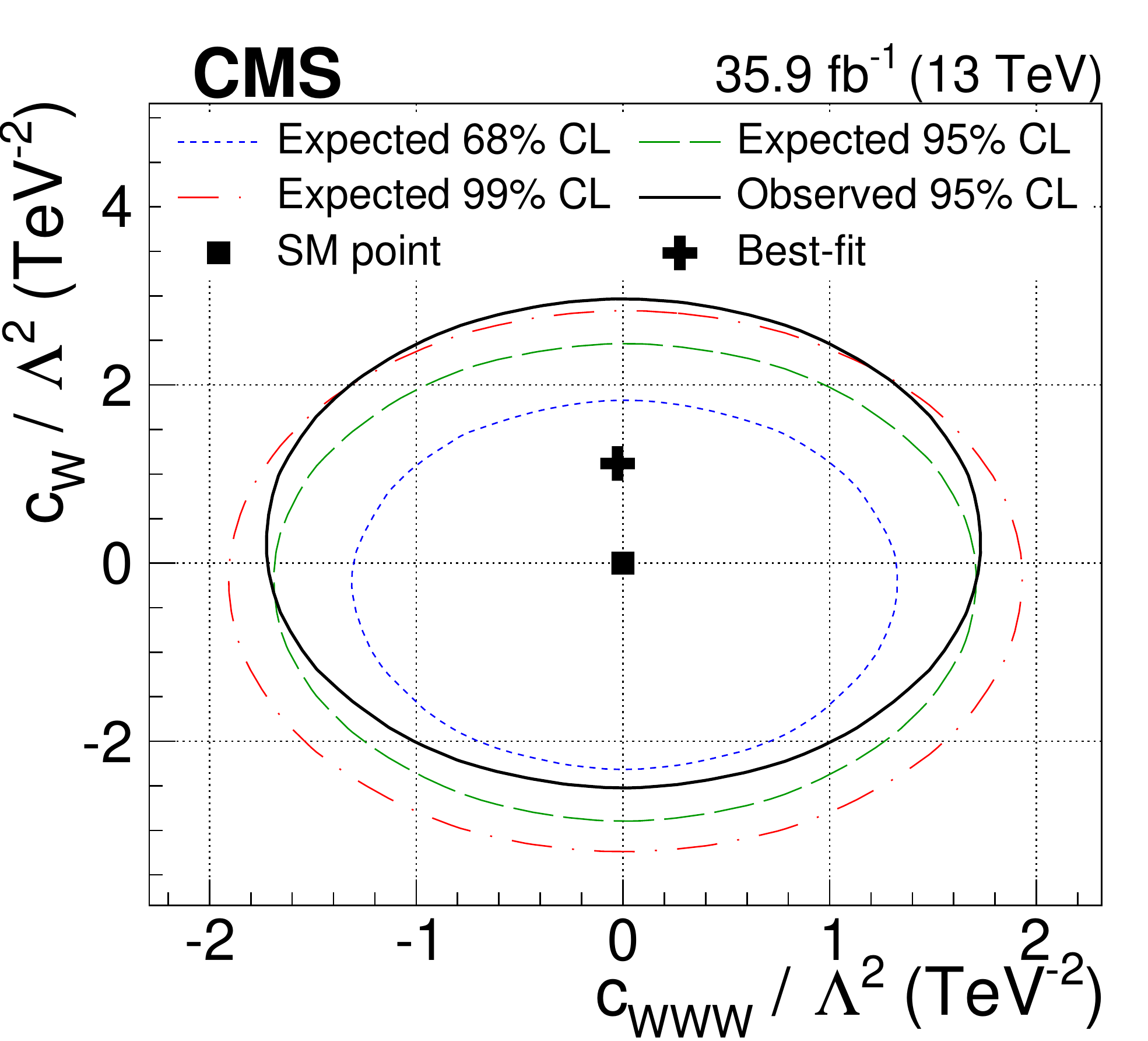}
\hfill
\includegraphics[width=0.32\textwidth]{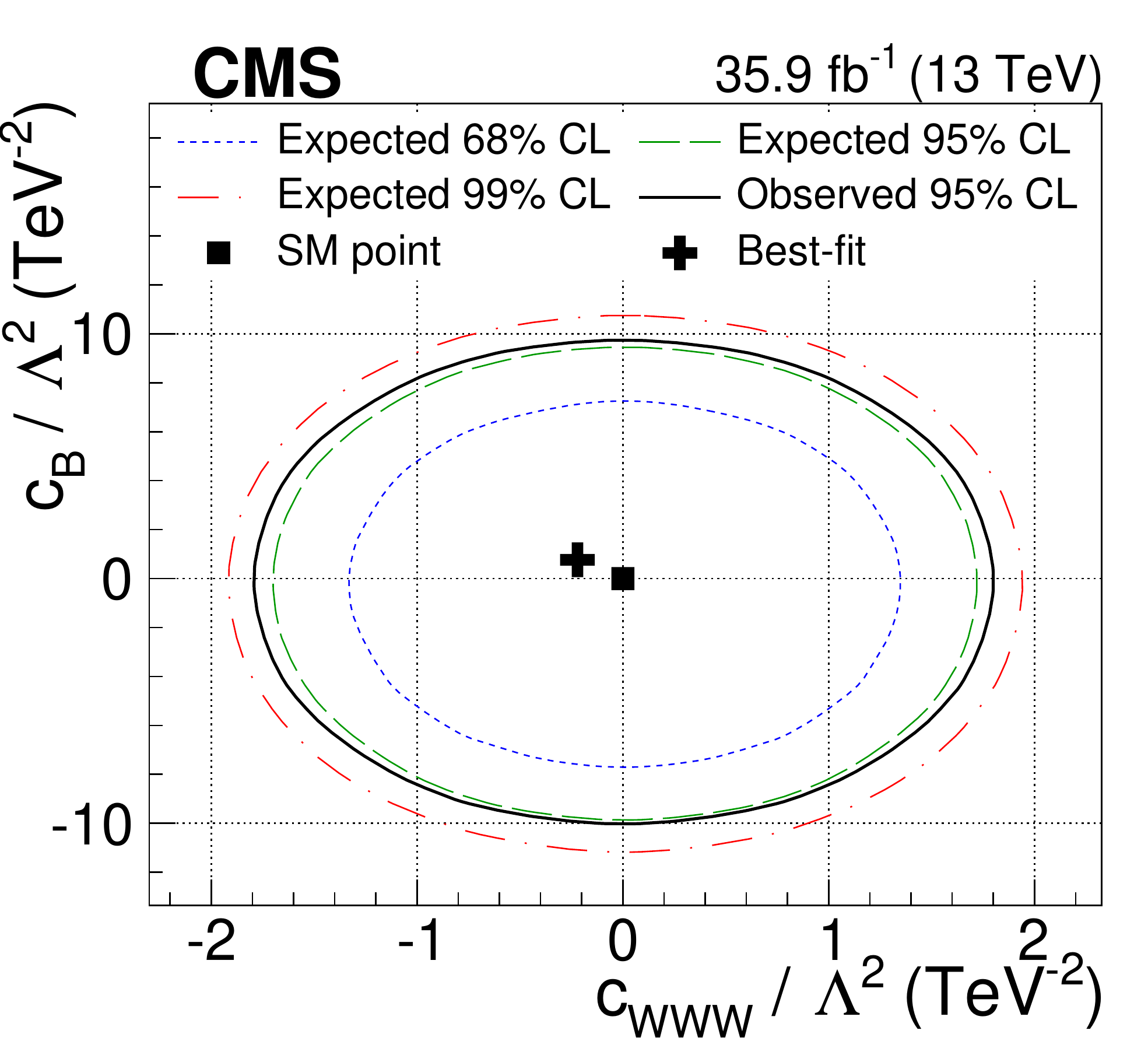}
\hfill
\includegraphics[width=0.32\textwidth]{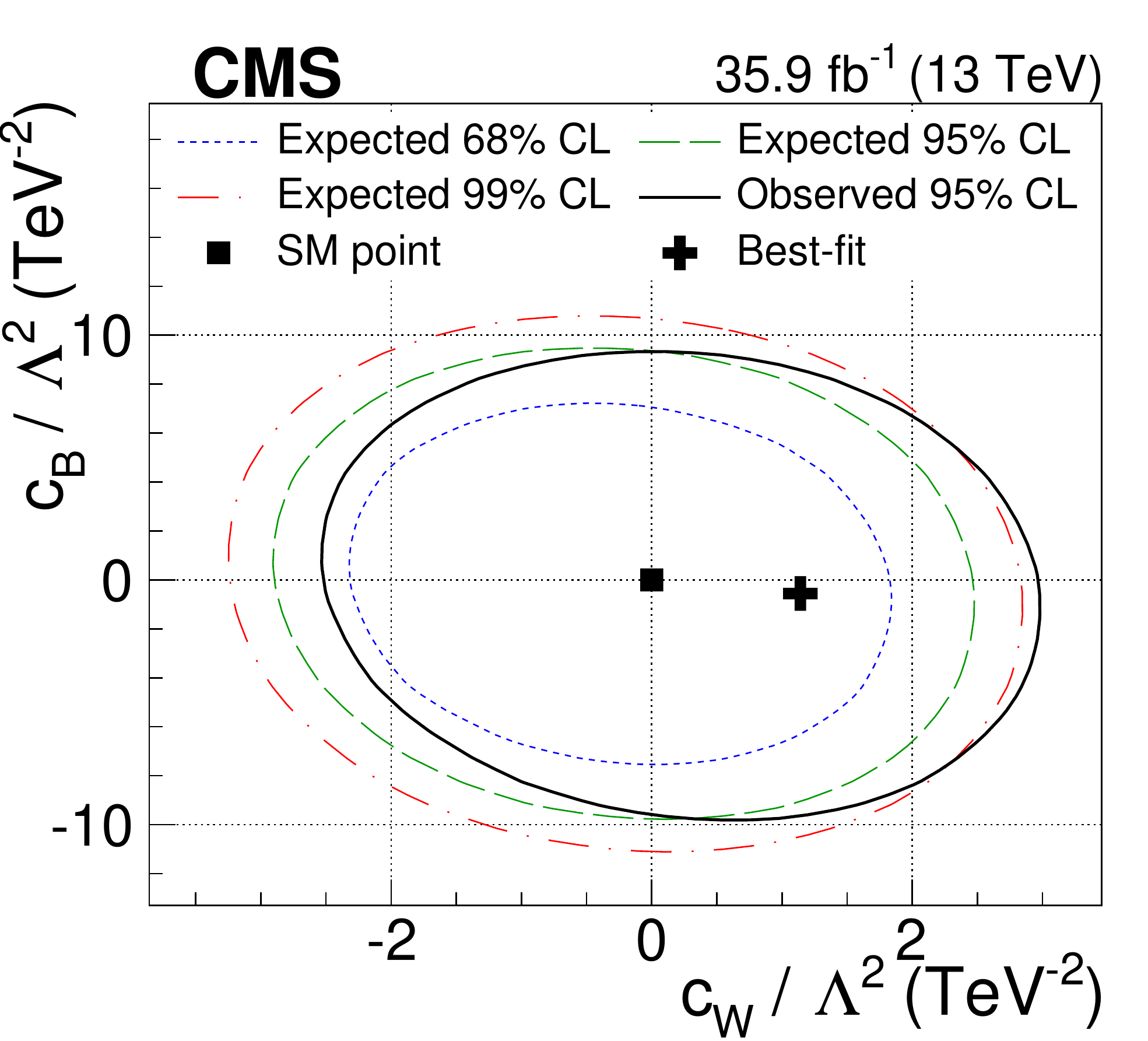}
\caption{Two-dimensional limits on the aTGC parameters in the EFT parametrization, for the combinations $\cwww/\Lambda^2$--$\cw/\Lambda^2$ (left), $\cwww/\Lambda^2$--$\cb/\Lambda^2$ (centre), and $\cw/\Lambda^2$--$\cb/\Lambda^2$ (right). Contours for the expected 95\% \CL are shown in dashed green, with the 68 and 99\% \CL contours shown in dotted blue and dot-dashed red, respectively. Contours for the observed 95\% \CL are shown in solid black. The black square markers represent the SM expectation, while the black crosses show the observed best-fit points.}
\label{fig:finallimit}
\end{figure}

\begin{figure}[h!]
\centering
\includegraphics[width=0.32\textwidth]{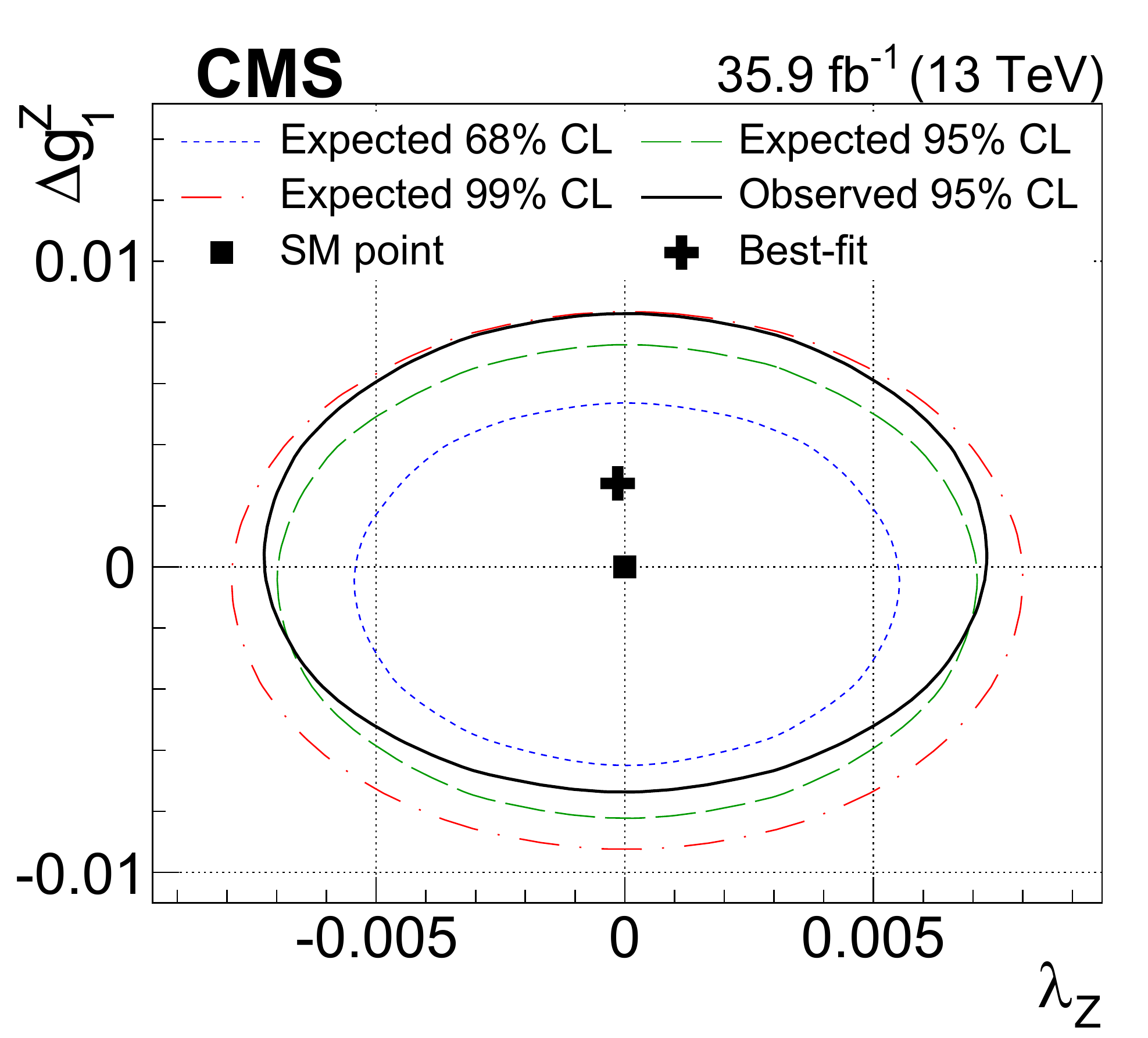}
\hfill
\includegraphics[width=0.32\textwidth]{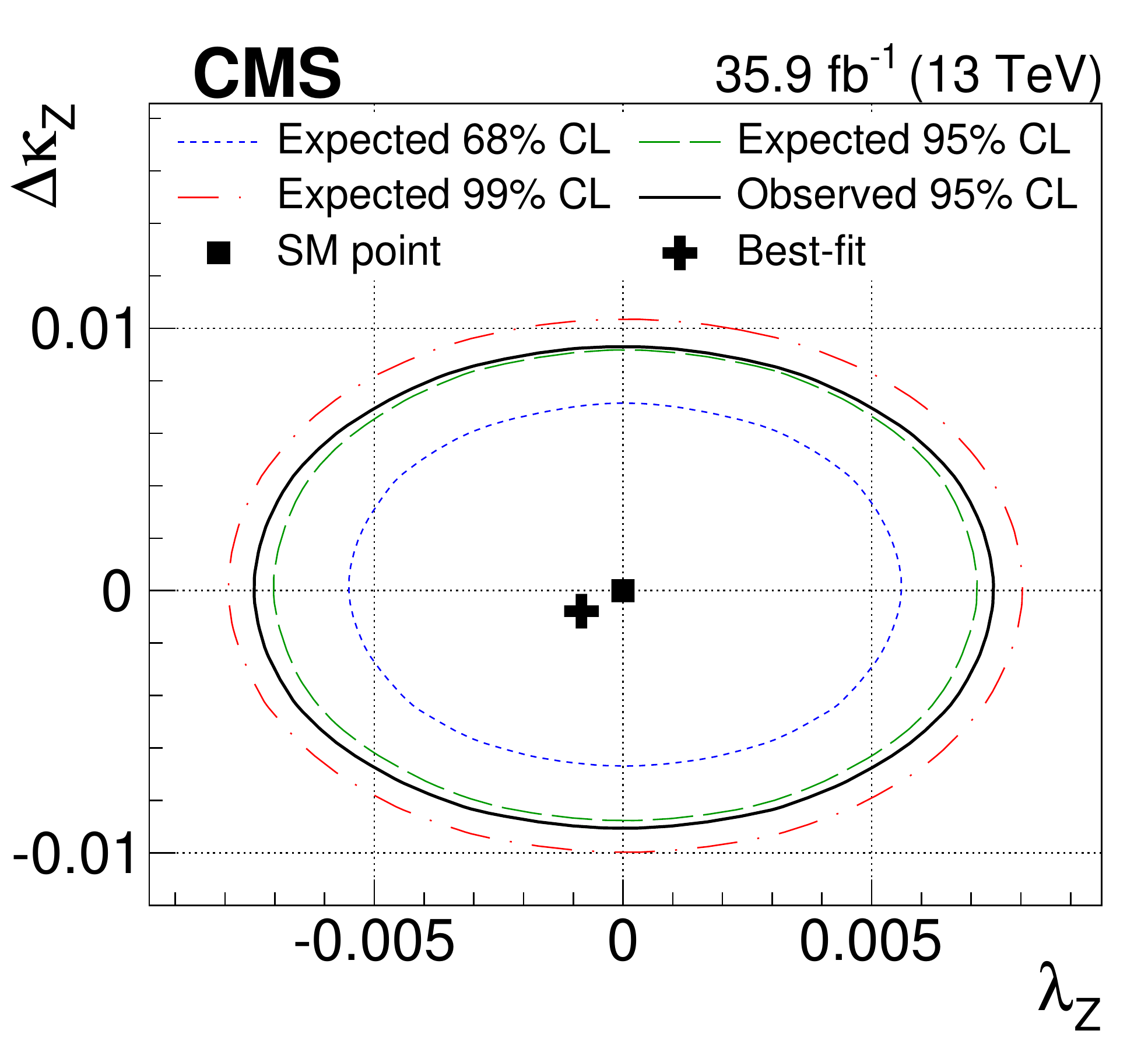}
\hfill
\includegraphics[width=0.32\textwidth]{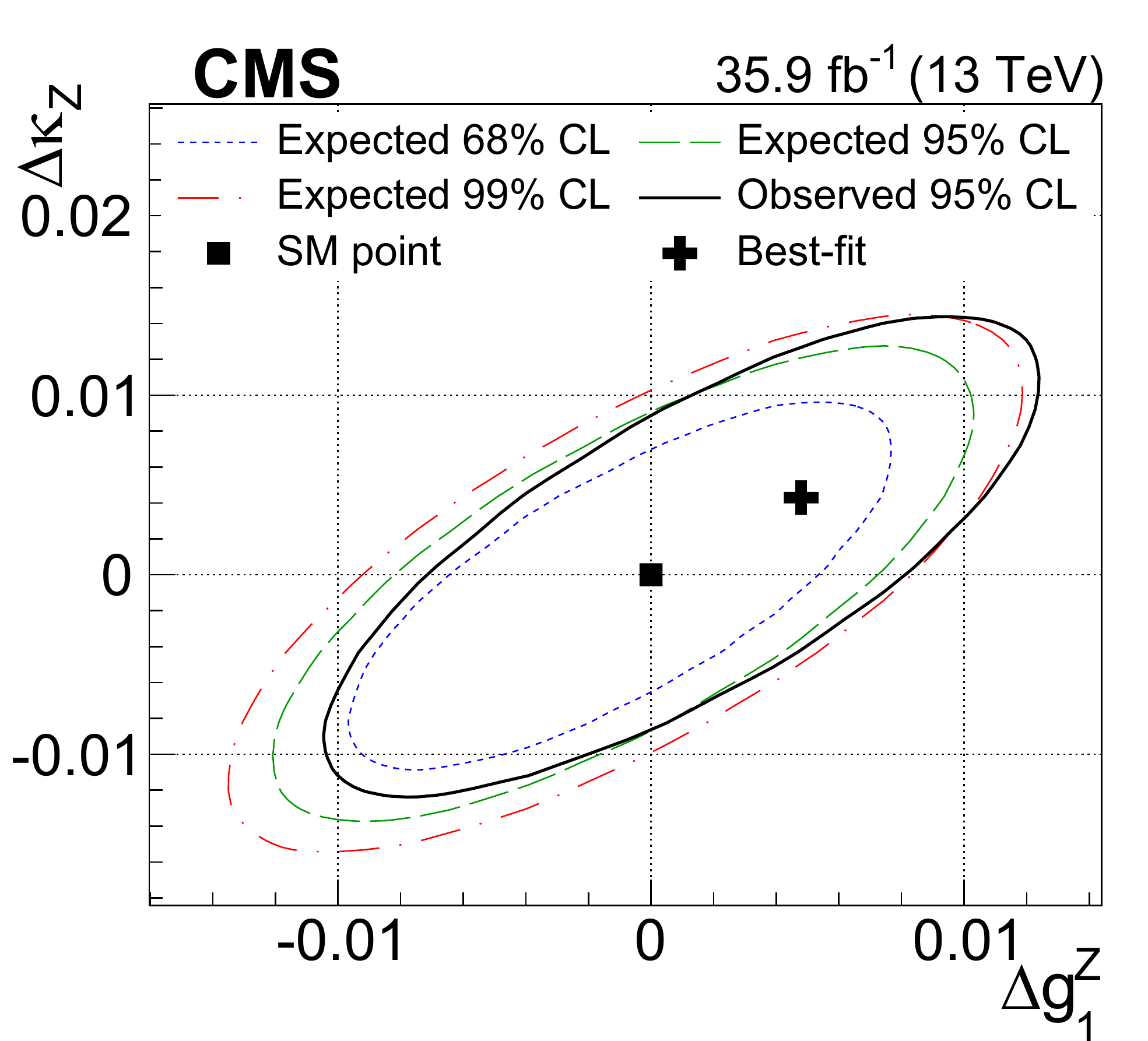}
\caption{Two-dimensional limits on the aTGC parameters in the LEP parametrization, for the combinations $\lambdaz$--$\deltagz$ (left), $\lambdaz$--$\deltakapz$ (centre), and $\deltagz$--$\deltakapz$ (right). Contours for the expected 95\% \CL are shown in dashed green, with the 68 and 99\% \CL contours shown in dotted blue and dot-dashed red, respectively. Contours for the observed 95\% \CL are shown in solid black. The black square markers represent the SM expectation, while the black crosses show the observed best-fit points.}
\label{fig:finallimitLEP}
\end{figure}

Two-dimensional expected and observed limits on pairwise combinations of the couplings, with the remaining coupling set to zero, are also derived, and the results shown in Fig.~\ref{fig:finallimit} for the EFT parametrization, and Fig.~\ref{fig:finallimitLEP} for the LEP parametrization.

\begin{table}[h!]
\centering
\topcaption{Expected and observed limits at 95\% \CL on single anomalous couplings, along with observed best-fit values, for both the EFT and LEP parametrizations. For each coupling, all other couplings are explicitly set to zero. Observed limits from collision data taken at a centre-of-mass energy of 8\TeV~\cite{Sirunyan:2017bey} are also quoted for comparison.}
\cmsTable{
\begin{tabular}{cccccc}
Parametrization & aTGC & Expected limit & Observed limit & Observed best-fit & $8\TeV$ observed limit \\
& \\[\dimexpr-\normalbaselineskip+2pt]
\hline
& \\[\dimexpr-\normalbaselineskip+2pt]
\multirow{3}{*}{EFT}
  &$\cwww/\Lambda^2$ $({\TeVns}^{-2})$ & [-1.44, 1.47]  & [-1.58, 1.59] & -0.26 & [-2.7, 2.7] \\
  &$\cw/\Lambda^2$ $({\TeVns}^{-2})$   & [-2.45, 2.08]  & [-2.00, 2.65] & 1.21  & [-2.0, 5.7] \\
  &$\cb/\Lambda^2$ $({\TeVns}^{-2})$   & [-8.38, 8.06]  & [-8.78, 8.54] & 1.07  & [-14, 17] \\ [\cmsTabSkip]

\multirow{3}{*}{LEP}
  &$\lambdaz$     & [-0.0060, 0.0061]  &  [-0.0065, 0.0066] & -0.0010 & [-0.011, 0.011] \\
  &$\deltagz$     & [-0.0070, 0.0061]  &  [-0.0061, 0.0074] & 0.0027  & [-0.009, 0.024 ] \\
  &$\deltakapz$   & [-0.0074, 0.0078]  &  [-0.0079, 0.0082] & -0.0010 & [-0.018, 0.013 ] \\
\hline
\end{tabular}
}
\label{tab:1dlim_vertex}
\end{table}

While the operators associated with $\cwww$ and $\cw$ induce contributions in similar proportions in both the $\WW$ and $\WZ$ signal regions, we expect the effects of the operator associated with $\cb$ to be much greater in the $\WW$ region compared to the $\WZ$ region. Consequently, there is little separation power in this analysis between $\cwww$ and $\cw$, with a similar limit derived on both couplings, and more separation power between $\cwww$/$\cw$ and $\cb$ in the case of nonzero coupling values.

\begin{figure}[h!]
\centering
\includegraphics[width=\textwidth]{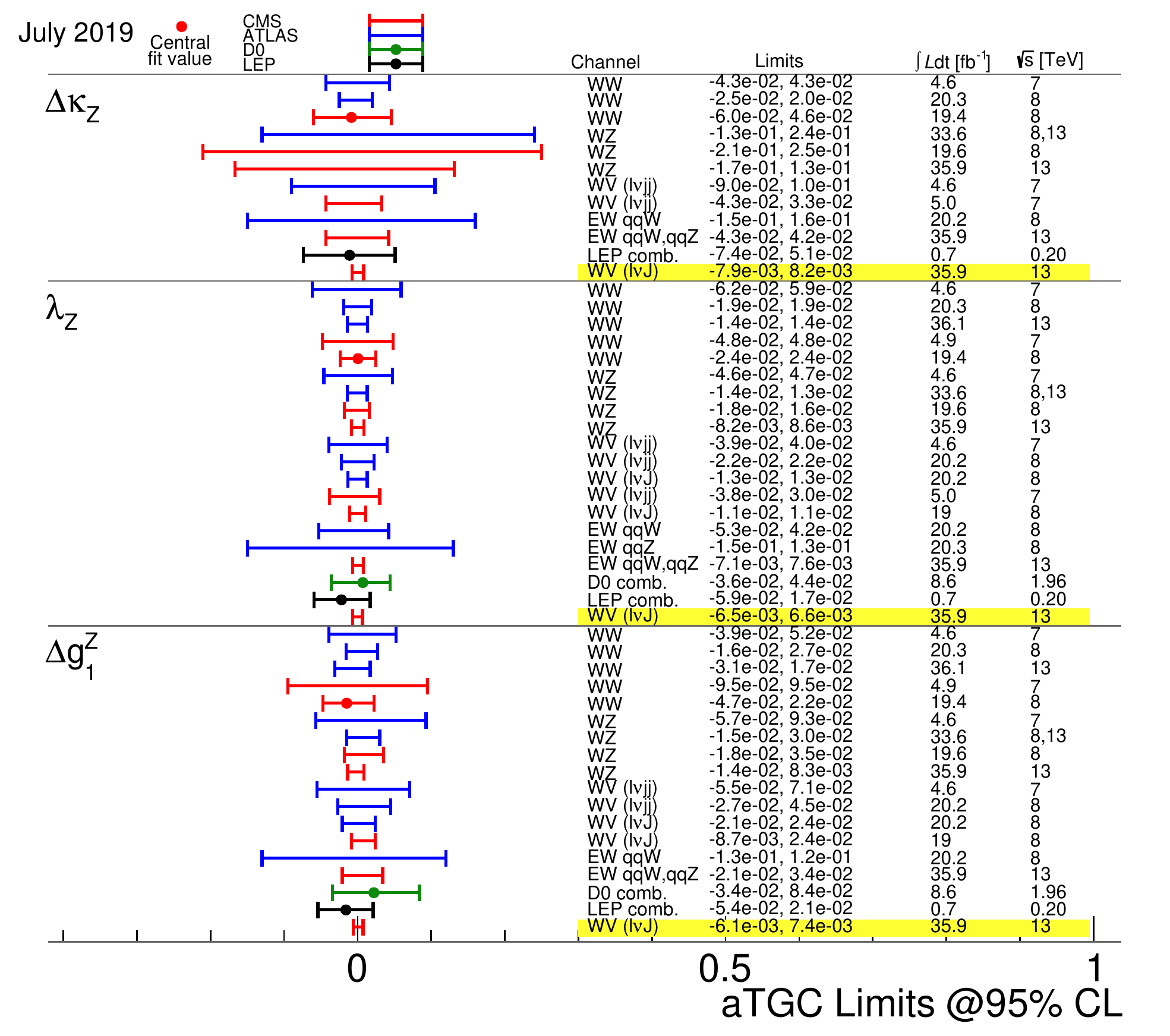}
\caption{Comparison of the observed limits on aTGC parameters in the LEP parametrization from different measurements. The highlighted rows represent the limits obtained from this measurement.}
\label{fig:aTGCsComparison}
\end{figure}

A comparison of limits derived in this analysis with those obtained by other analyses performed at the LEP~\cite{Schael:2013ita}, \DZERO~\cite{Abazov:2012ze}, CMS~\cite{Chatrchyan:2013yaa,Khachatryan:2016poo,Sirunyan:2019bez,Chatrchyan:2012bd,Sirunyan:2017bey,Sirunyan:2017jej,Sirunyan:2019dyi}, and ATLAS~\cite{ATLAS:2012mec,Aad:2016wpd,Aaboud:2019nkz,Aad:2012twa,Aad:2016ett,Aad:2014mda,Aaboud:2017cgf,Aad:2014dta,Aaboud:2017fye,ATLAS-CONF-2016-043} experiments is shown in Fig.~\ref{fig:aTGCsComparison}.
Limits that were set on $\lambda_{\Pg}$ and $\deltakapg$ have been converted to limits on $\lambdaz$ and $\deltakapz$, respectively, using the relationships in Ref.~\cite{Degrande:2012wf}.
The limits derived in this analysis are the strictest bounds on all three parameters to date, improving upon the complementary all-leptonic searches also performed using collision data recorded at a centre-of-mass energy of 13\TeV by the ATLAS~\cite{Aaboud:2019nkz,ATLAS-CONF-2016-043} and CMS~\cite{Sirunyan:2019bez} Collaborations.
There is an especially significant improvement in the measured limit on $\deltakapz$ over any previous measurement.

\section{Summary}
\label{sec:summary}

A measurement of limits on anomalous triple gauge coupling parameters in terms of dimension-six effective field theory operators has been presented.
It uses events where two vector bosons are produced, with one decaying leptonically and the other hadronically to a single, massive, large-radius jet.
Results are based on data recorded in proton-proton collisions at $\sqrt{s}=13 \TeV$ with the CMS detector at the CERN LHC in 2016, corresponding to an integrated luminosity of 35.9\fbinv.
Limits are presented both in terms of the $\cwww$, $\cw$, and $\cb$ parameters (scaled by an overall new physics energy scale $\Lambda$) in the effective field theory parametrization, and the $\lambdaz$, $\deltagz$, and $\deltakapz$ parameters in the LEP parametrization.
For each parametrization, limits are set at 95\% confidence level on individual parameters, as well as on pairwise combinations of parameters.
Limits on individual parameters in the effective field theory parametrization are determined to be $-1.58 < \cwww/\Lambda^2 < 1.59\TeV^{-2}$, $-2.00 < \cw/\Lambda^2 < 2.65\TeV^{-2}$, and $-8.78 < \cb/\Lambda^2 < 8.54\TeV^{-2}$, in agreement with standard model expectations of zero for each parameter.
These are the strictest bounds on these parameters to date.

\begin{acknowledgments}
We congratulate our colleagues in the CERN accelerator departments for the excellent performance of the LHC and thank the technical and administrative staffs at CERN and at other CMS institutes for their contributions to the success of the CMS effort. In addition, we gratefully acknowledge the computing centres and personnel of the Worldwide LHC Computing Grid for delivering so effectively the computing infrastructure essential to our analyses. Finally, we acknowledge the enduring support for the construction and operation of the LHC and the CMS detector provided by the following funding agencies: BMBWF and FWF (Austria); FNRS and FWO (Belgium); CNPq, CAPES, FAPERJ, FAPERGS, and FAPESP (Brazil); MES (Bulgaria); CERN; CAS, MoST, and NSFC (China); COLCIENCIAS (Colombia); MSES and CSF (Croatia); RPF (Cyprus); SENESCYT (Ecuador); MoER, ERC IUT, PUT and ERDF (Estonia); Academy of Finland, MEC, and HIP (Finland); CEA and CNRS/IN2P3 (France); BMBF, DFG, and HGF (Germany); GSRT (Greece); NKFIA (Hungary); DAE and DST (India); IPM (Iran); SFI (Ireland); INFN (Italy); MSIP and NRF (Republic of Korea); MES (Latvia); LAS (Lithuania); MOE and UM (Malaysia); BUAP, CINVESTAV, CONACYT, LNS, SEP, and UASLP-FAI (Mexico); MOS (Montenegro); MBIE (New Zealand); PAEC (Pakistan); MSHE and NSC (Poland); FCT (Portugal); JINR (Dubna); MON, RosAtom, RAS, RFBR, and NRC KI (Russia); MESTD (Serbia); SEIDI, CPAN, PCTI, and FEDER (Spain); MOSTR (Sri Lanka); Swiss Funding Agencies (Switzerland); MST (Taipei); ThEPCenter, IPST, STAR, and NSTDA (Thailand); TUBITAK and TAEK (Turkey); NASU and SFFR (Ukraine); STFC (United Kingdom); DOE and NSF (USA).

\hyphenation{Rachada-pisek} Individuals have received support from the Marie-Curie programme and the European Research Council and Horizon 2020 Grant, contract Nos.\ 675440, 752730, and 765710 (European Union); the Leventis Foundation; the A.P.\ Sloan Foundation; the Alexander von Humboldt Foundation; the Belgian Federal Science Policy Office; the Fonds pour la Formation \`a la Recherche dans l'Industrie et dans l'Agriculture (FRIA-Belgium); the Agentschap voor Innovatie door Wetenschap en Technologie (IWT-Belgium); the F.R.S.-FNRS and FWO (Belgium) under the ``Excellence of Science -- EOS" -- be.h project n.\ 30820817; the Beijing Municipal Science \& Technology Commission, No. Z181100004218003; the Ministry of Education, Youth and Sports (MEYS) of the Czech Republic; the Lend\"ulet (``Momentum") Programme and the J\'anos Bolyai Research Scholarship of the Hungarian Academy of Sciences, the New National Excellence Program \'UNKP, the NKFIA research grants 123842, 123959, 124845, 124850, 125105, 128713, 128786, and 129058 (Hungary); the Council of Science and Industrial Research, India; the HOMING PLUS programme of the Foundation for Polish Science, cofinanced from European Union, Regional Development Fund, the Mobility Plus programme of the Ministry of Science and Higher Education, the National Science Center (Poland), contracts Harmonia 2014/14/M/ST2/00428, Opus 2014/13/B/ST2/02543, 2014/15/B/ST2/03998, and 2015/19/B/ST2/02861, Sonata-bis 2012/07/E/ST2/01406; the National Priorities Research Program by Qatar National Research Fund; the Ministry of Science and Education, grant no. 3.2989.2017 (Russia); the Programa Estatal de Fomento de la Investigaci{\'o}n Cient{\'i}fica y T{\'e}cnica de Excelencia Mar\'{\i}a de Maeztu, grant MDM-2015-0509 and the Programa Severo Ochoa del Principado de Asturias; the Thalis and Aristeia programmes cofinanced by EU-ESF and the Greek NSRF; the Rachadapisek Sompot Fund for Postdoctoral Fellowship, Chulalongkorn University and the Chulalongkorn Academic into Its 2nd Century Project Advancement Project (Thailand); the Welch Foundation, contract C-1845; and the Weston Havens Foundation (USA). \end{acknowledgments}

\bibliography{auto_generated}
\cleardoublepage \appendix\section{The CMS Collaboration \label{app:collab}}\begin{sloppypar}\hyphenpenalty=5000\widowpenalty=500\clubpenalty=5000\vskip\cmsinstskip
\textbf{Yerevan Physics Institute, Yerevan, Armenia}\\*[0pt]
A.M.~Sirunyan$^{\textrm{\dag}}$, A.~Tumasyan
\vskip\cmsinstskip
\textbf{Institut f\"{u}r Hochenergiephysik, Wien, Austria}\\*[0pt]
W.~Adam, F.~Ambrogi, T.~Bergauer, J.~Brandstetter, M.~Dragicevic, J.~Er\"{o}, A.~Escalante~Del~Valle, M.~Flechl, R.~Fr\"{u}hwirth\cmsAuthorMark{1}, M.~Jeitler\cmsAuthorMark{1}, N.~Krammer, I.~Kr\"{a}tschmer, D.~Liko, T.~Madlener, I.~Mikulec, N.~Rad, J.~Schieck\cmsAuthorMark{1}, R.~Sch\"{o}fbeck, M.~Spanring, D.~Spitzbart, W.~Waltenberger, C.-E.~Wulz\cmsAuthorMark{1}, M.~Zarucki
% ...
\vskip\cmsinstskip
\dag: Deceased\\
1:  Also at Vienna University of Technology, Vienna, Austria\\
% ...
\end{sloppypar}
\end{document}